\documentclass[a4paper,11pt]{article}
\pdfoutput=1

\usepackage{jheppub}
\usepackage{graphicx}
\usepackage{amsmath}
\usepackage{xspace}
\usepackage{slashed}
\usepackage{color}

\newcommand{\eq}[1]{eq.~\eqref{eq:#1}}
\newcommand{\eqs}[2]{eqs.~\eqref{eq:#1} and \eqref{eq:#2}}
\newcommand{\eqsc}[2]{eqs.~\eqref{eq:#1},~\eqref{eq:#2}}
\newcommand{\eqss}[2]{eqs.~\eqref{eq:#1}-\eqref{eq:#2}}
\renewcommand{\sec}[1]{section~\ref{sec:#1}}

\newcommand{\subsec}[1]{section~\ref{subsec:#1}}
\newcommand{\subsecs}[2]{sections~\ref{subsec:#1} and \ref{subsec:#2}}
\newcommand{\subsubsec}[1]{section~\ref{subsubsec:#1}}

\newcommand{\app}[1]{appendix~\ref{app:#1}}
\newcommand{\fig}[1]{figure~\ref{fig:#1}}
\newcommand{\figs}[2]{figures~\ref{fig:#1} and \ref{fig:#2}}
\newcommand{\Tab}[1]{table~\ref{tab:#1}}
\newcommand{\mycites}[1]{refs.~\cite{#1}}
\newcommand{\mycite}[1]{ref.~\cite{#1}}

\newcommand{\abs}[1]{\lvert#1\rvert}

\newcommand{\ordo}[1]{\mathcal{O}\left(#1\right)}

\newcommand{\ord}{{\mathcal O}}
\newcommand{\eps}{\epsilon}

\newcommand{\bra}[1]{\left\langle #1\right|}
\newcommand{\ket}[1]{\left| #1\right\rangle}

\newcommand{\df}{\mathrm{d}}

\newcommand{\tr}{\mathrm{Tr}}

\newcommand{\e}{\epsilon}

\newcommand{\nn}{\nonumber}
\newcommand{\bb}{\mathbf{b}}

\newcommand{\cusp}{\mathrm{cusp}}

\newcommand{\bare}{\mathrm{b}}

\newcommand{\one}{{(1)}}
\newcommand{\two}{{(2)}}

\newcommand{\lqcd}{\Lambda_\mathrm{QCD}}
\newcommand{\MSbar}{$\overline{\text{MS}}$\xspace}

\newcommand{\asmu}{\ensuremath{\left(\frac{\alpha_s(\mu)}{4\pi}\right)}}
\newcommand{\pt}{\mathbf{p}_\perp}

\newcommand{\plus}[1]{\mathcal{L}_{#1}^T}
\newcommand{\Log}{\ln\frac{\mu}{\nu}}
\newcommand{\Logs}{\ln^2\frac{\mu}{\nu}}
\newcommand{\ptdelta}{\delta^{(2)}(\pt)}

\newcommand{\bn}{\bar{n}}

\allowdisplaybreaks[3]

\title{Rapidity renormalized TMD soft and beam functions at two loops}

\author[a]{Thomas L\"{u}bbert,}
\author[b,c]{Joel Oredsson,}
\author[b,d]{Maximilian Stahlhofen}

\emailAdd{thomas.luebbert@desy.de}
\emailAdd{joel.oredsson@thep.lu.se}
\emailAdd{mastahlh@uni-mainz.de}

\affiliation[a]{II. Institute for Theoretical Physics, University of Hamburg, 22761 Hamburg, Germany}
\affiliation[b]{Theory Group, Deutsches Elektronen-Synchrotron (DESY), Notkestra\ss e 85, D-22607 Hamburg, Germany}
\affiliation[c]{Department of Astronomy and Theoretical Physics, Lund University, S\"{o}lvegatan 14A 223 62 Lund, Sweden}
\affiliation[d]{PRISMA Cluster of Excellence, Institute of Physics, Johannes Gutenberg University, Staudingerweg 7, 55128 Mainz, Germany}

\abstract{
We compute the transverse momentum dependent (TMD) soft function 
for the production of a color-neutral final state at the LHC within the rapidity renormalization group (RRG) framework to next-to-next-to-leading order (NNLO).
We use this result to extract the universal renormalized TMD beam functions (aka TMDPDFs) 
in the same scheme and at the same order from known results in another scheme.
We derive recurrence relations for the logarithmic structure of the soft and beam functions, which we use to cross check our calculation.
We also explicitly confirm the non-Abelian exponentiation of the TMD soft function in the RRG framework at two loops.
Our results provide the ingredients for resummed predictions of $\pt$-differential cross sections at NNLL$'$ in the RRG formalism. 
The RRG provides a systematic framework to resum large (rapidity) logarithms 
through (R)RG evolution and to assess the associated perturbative uncertainties.
}


\begin{document}

\preprint{
\begin{flushright}
DESY 16-024\\
MITP 16-017
\end{flushright}
}

\maketitle

\section{Introduction}
\label{sec:intro}

The transverse momentum ($\pt$) distribution of a heavy color-neutral final state $L$ in the process
\begin{align}
p + p \to L + X\,,
\label{eq:pptoLX}
\end{align}
is an important and precisely measurable observable at the LHC. Here $X$ denotes an arbitrary unresolved final state.
Prominent examples for such processes are the 
Drell-Yan-type production of electroweak gauge bosons and Higgs bosons via quark-antiquark annihilation and gluon fusion (given the bosons decay non-hadronically).
For large transverse momenta $\pt^2 \sim M^2$ with $M$ being the invariant mass of $L$, fixed-order (FO) QCD calculations at sufficiently high order provide an accurate description of the $\pt$-spectrum.
In the region where $\pt^2 \ll M^2$, the distribution is peaked and the resummation of (Sudakov) logarithms $\propto \ln^n(\pt^2/M^2)$ is crucial for reliable theoretical predictions.

The traditional approach to resum such logarithms in QCD is based on the factorization theorem devised by Collins, Soper, and Sterman (CSS) in \mycite{Collins:1984kg}. In this framework the resummation has reached the next-to-next-to-leading-logarithmic (NNLL) level~\cite{deFlorian:2000pr,deFlorian:2001zd,Bozzi:2005wk,Catani:2010pd,deFlorian:2011xf,Catani:2011kr, Catani:2012qa,Catani:2013tia}.
In recent years, the factorization for $\pt$-differential cross sections has been revisited~\cite{Becher:2010tm,GarciaEchevarria:2011rb,Chiu:2012ir}
using Soft Collinear Effective Theory (SCET)~\cite{Bauer:2000ew,Bauer:2000yr,Bauer:2001ct,Bauer:2001yt,Bauer:2002aj,Beneke:2002ph}. 
Also a number of SCET predictions with NNLL resummation for Drell-Yan-like processes are available by now~\cite{Becher:2011xn,Becher:2012yn,Neill:2015roa,Echevarria:2015uaa}. The SCET approach typically allows a direct control of the perturbative uncertainties by independent variations of the various unphysical factorization/renormalization scales in the problem, both in momentum and position (impact parameter) space. In principle also subleading power corrections of relative $\ord(\pt^2/M^2)$ to the factorization theorem are systematically calculable in SCET.

Throughout this paper we will focus on the perturbatively accessible resummation region of the $\pt$-spectrum, i.e.\ where $\lqcd^2 \ll \pt^2 \ll M^2$.
The leading order SCET factorization theorem generically has the schematic form
\begin{align}
\frac{\df\sigma}{\df\pt^2} = H \times \Big[S \otimes B_a \otimes B_b \Big](\pt) \,.
\label{eq:SchemFactTh}
\end{align}
The process-dependent (but $\pt$-independent) hard function $H$ contains all purely virtual QCD corrections to the partonic production of $L$.
The hard function multiplies a convolution of 
one TMD soft function $S(\pt)$
and
two TMD beam functions $B_{a,b}(\pt)$ -- one for each beam. The latter are also called TMD parton distribution functions (TMDPDFs) in the literature.\footnote{The name TMDPDF goes back to the traditional QCD factorization formalism. We prefer to call it the TMD beam function, as in SCET it is conceptually on the same footing as the virtuality-dependent~\cite{Stewart:2009yx,Stewart:2010qs,Gaunt:2014xga,Gaunt:2014cfa} or fully-differential (unintegrated) beam function~\cite{Mantry:2009qz,Jain:2011iu,Gaunt:2014xxa} (although the latter are SCET$_{\rm I}$ objects).}
The beam and soft functions describe the collinear and soft radiation off the incoming hard partons, respectively.
The beam functions are process-independent (universal). The soft function (for a given partonic channel) is universal within the class of processes in \eq{pptoLX}, i.e.\ independent of $L$. Up to two loops it is the same for Drell-Yan like processes, $e^+ e^- \to$ dijets and deep inelastic scattering~\cite{Kang:2015moa}.

It is well known that the factorization of the $\pt$-differential cross section into separate soft and collinear sectors gives rise to so-called rapidity divergences, see e.g.\ \mycite{Collins:2011zzd,Chiu:2012ir}. They arise from loop/phase space (PS) momentum regions, where one light-cone component, say $k^+$, approaches infinity or zero, while the product $k^+ k^- \sim k^2$ is kept fixed.
This typically happens in soft and beam functions, when the corresponding soft and collinear modes relevant to describe the process have virtualities of the same order of magnitude. The production of a final state $L$ with measured transverse momentum $\pt$, which recoils against the soft as well as the collinear real radiation, is a classic example for such a process. The appropriate effective field theory (EFT) framework for this kinematic situation is SCET$_{\rm II}$.
By contrast, the virtuality scales of (ultra)soft and collinear degrees of freedom in SCET$_{\rm I}$ are far separated. Soft and collinear functions in SCET$_{\rm I}$ factorization theorems, e.g.\ for thrust or $N$-jettiness, are therefore free from rapidity divergences.
 
Unlike divergences that are related to Lorentz invariant products of loop/PS momenta approaching ultraviolet (UV) or infrared (IR) limits, rapidity divergences are not regulated by dimensional regularization.
In the context of SCET several appropriate regulators have been proposed in the literature, see e.g.\ \mycites{Chiu:2009yx,Becher:2011dz,Chiu:2011qc,Chiu:2012ir,Echevarria:2015usa}.
Using these in addition to dimensional regularization ($d=4-2\eps$) the bare beam and soft functions can be defined separately.
Once they are combined as in \eq{SchemFactTh} the dependence on the rapidity regulator cancels out order by order in FO perturbation theory leaving only overall $1/\eps^n$ UV poles, which eventually cancel in the product with the bare hard function.\footnote{The 'bare' hard function we refer to here consists of the IR divergent purely virtual QCD corrections to the (partonic) hard process (plus Born level) in dimensional regularization.}
In the renormalized version of the factorization theorem they are removed by respective counterterms of the beam and soft functions.

In analogy to the usual UV renormalization procedure also the rapidity divergences can be removed (after zero-bin subtractions~\cite{Manohar:2006nz,Chiu:2009yx}) from the beam and soft functions leading to finite, but renormalization scheme dependent, (rapidity-)renormalized results. 
Just like the rapidity regulator the scheme dependence cancels out in FO cross sections.
Different subtraction schemes have been used in the literature, cf. \mycites{Becher:2010tm,GarciaEchevarria:2011rb,Chiu:2012ir}.\footnote{Here we adopt the rapidity renormalization group point of view according to \mycites{Chiu:2011qc,Chiu:2012ir} and interpret \mycites{Becher:2010tm,GarciaEchevarria:2011rb} in this context.}
In this paper we will employ a minimal subtraction scheme together with the rapidity regulator $\eta$ as proposed in \mycites{Chiu:2011qc,Chiu:2012ir}.
This scheme allows to straightforwardly set up a rapidity renormalization group (RRG) in addition to the usual renormalization group (RG).
That is to say, the renormalized beam and soft functions separately fulfill (entangled) RG equations (RGEs) and rapidity RGEs (RRGEs) describing the dependence on the UV and rapidity renormalization scales $\mu_{B,S}$ and $\nu_{B,S}$, respectively. These can be used to resum the complete set of logarithms of the ratio $\pt^2/M^2$ as shown in detail in \mycite{Chiu:2012ir}.

In order to improve the precision of the resummed SCET predictions for the $\pt$-differential production of $L$ beyond NNLL, i.e.\ NNLL$'$ (and eventually N$^3$LL)\footnote{For details about the prime counting (NLL$'$, NNLL$'$, \ldots) see e.g.\ \mycites{Abbate:2010xh,Berger:2010xi,Stewart:2013faa,Almeida:2014uva}.},
the FO expressions for the beam and soft function are required at NNLO.
At present the complete set of NNLO results is only available in the scheme of \mycite{Becher:2010tm}, see \mycites{Gehrmann:2012ze,Gehrmann:2014yya}.
In the scheme of \mycite{GarciaEchevarria:2011rb} only the soft function has recently been computed at NNLO~\cite{Echevarria:2015byo}.
The aim of the present paper is to determine the NNLO beam and soft functions in the scheme of \mycites{Chiu:2011qc,Chiu:2012ir}.
We note that the universal part of the (hard-collinear) factorized cross section has first been computed to NNLO in the traditional QCD approach~\cite{Catani:2011kr,Catani:2012qa}.

The RRG framework of \mycites{Chiu:2011qc,Chiu:2012ir} has the advantage that the perturbative uncertainties related to the dependence on the (unphysical) hard, beam and soft (rapidity) renormalization scales ($\mu_H$, $\mu_B$, $\mu_S$, $\nu_B$, $\nu_S$) can be assessed directly and independently.
This is done 
using so-called profile functions in order to independently vary the scales in the resummed result for the cross section, while ensuring a smooth transition to the FO regime at large $\pt$, see e.g.\ \mycites{Stewart:2013faa,Neill:2015roa}. 
In any case we believe that given the importance of the observable it will be beneficial to eventually have precise theoretical predictions in different resummation frameworks. This will provide important cross checks among the different approaches and also contribute to a better theoretical error estimate.

The outline of this paper is as follows.
In \sec{setup}, we will briefly review the SCET definition of TMD soft and beam functions, their regularization and renormalization following \mycites{Chiu:2011qc,Chiu:2012ir}. 
Their all order logarithmic structure is discussed in \sec{logstruc}.
In \sec{calculation}, we calculate the TMD soft function through NNLO 
and explicitly check its exponentiation properties and logarithmic structure.
We discuss the relation of our results to results in the literature
-- most importantly to \mycites{Gehrmann:2012ze,Gehrmann:2014yya} -- in \sec{comparison}.
This relation together with our result for the soft function allows us to extract the NNLO TMD beam functions in the RRG scheme of \mycites{Chiu:2011qc,Chiu:2012ir}. 
The expressions for the latter are presented in \sec{beam}.
Like all our results they are given in both momentum and position space.
We conclude in \sec{conclusions} and collect some additional technical details in the appendices.

\section{SCET framework}
\label{sec:setup}

In this paper we consider the
QCD corrections to the soft and beam functions in the SCET factorization theorem for the $\pt$-differential cross section of the process in \eq{pptoLX}, where $X$ now represents the inclusive hadronic final state.

We work in light-cone coordinates w.r.t.\ the light-like vectors $n^\mu=(1,0,0,1)$ and $\bn^\mu=(1,0,0,-1)$ along the beam (z-) axis. 
An arbitrary four-momentum vector is then expressed as
\begin{align}
p^\mu = p^- \frac{n^\mu}{2} + p^+ \frac{\bn^\mu}{2} + p_\perp^\mu\,,
\end{align}
where $p^+=n\cdot p$ and $p^-=\bn\cdot p$, or short $p^\mu = (p^+,p^-,p_\perp)$. Euclidean (two-)vectors are denoted in boldface, $\pt^2=-p_\perp^2$.
The partons initiating the hard interaction carry the (large) longitudinal momentum $\omega_\pm=Qe^{\pm Y}$, where the hard scale $Q=M [1+\ord(\pt^2/M^2)]$ and $Y$ is the rapidity of the color singlet state $L$.
The total center of mass energy of the colliding protons is $\sqrt{s}$.

For quarks as the incoming partons we write the factorized differential cross section as%
\footnote{To keep the notation simple, we do not write the quark 'charge' weighted sum over quark flavors here.}
\begin{align}
\frac{\df^3\sigma}{\df\mathbf{p}_{\perp}^2\df Q^2 \df Y}&= H_b(Q^2)\int \df^2\mathbf{p}_{1\perp}\df^2\mathbf{p}_{2\perp}\df^2\mathbf{p}_{s\perp}
\,\delta(\mathbf{p}_{\perp}^2-|\mathbf{p}_{1\perp}
+\mathbf{p}_{2\perp}+\mathbf{p}_{s\perp}|^2)\nn\\
&\times B_{q}^\bare(\omega_-/\sqrt{s},\mathbf{p}_{1\perp})
\,B_{\bar{q}}^\bare(\omega_+/\sqrt{s},\mathbf{p}_{2\perp})
\,S^q_\bare(\mathbf{p}_{s\perp})\,.
\label{eq:factq}
\end{align}
The index $b$ indicates bare quantities.
Similarly for incoming gluons we have%
\footnote{This holds for the production of spin 0 particles, as the Higgs boson for example. In the most general case, the beam functions are contracted with a non-diagonal Lorentz 4-tensor. }
\begin{align}
\frac{\df^3\sigma}{\df\mathbf{p}_{\perp}^2\df Q^2\df Y}&=H_b(Q^2)\int \df^2\mathbf{p}_{1_\perp}\df^2\mathbf{p}_{2\perp}\df^2\mathbf{p}_{s\perp}
\,\delta(\mathbf{p}_{\perp}^2-|\mathbf{p}_{1\perp}+\mathbf{p}_{2\perp}+\mathbf{p}_{s\perp}|^2)\nn\\
&\times B_g^{\bare,\mu\nu}(\omega_-/\sqrt{s},\mathbf{p}_{1\perp})
\,B_{g,\mu\nu}^\bare(\omega_+/\sqrt{s},\mathbf{p}_{2\perp})
\,S^g_\bare(\mathbf{p}_{s\perp})\,.
\label{eq:factg}
\end{align}
The 'bare' hard function $H_b$ in SCET can be expressed in terms of (the absolute square of) a hard current matching coefficient and the combined counterterms of beam and soft functions.
It is process dependent and will not be discussed any further in this paper.
The factorized cross section in terms of renormalized hard, beam and soft functions takes the same form as \eqs{factq}{factg}.
We simply replace bare by renormalized quantities. 

The bare soft functions are vacuum matrix elements of soft Wilson lines and are defined in momentum space as
\begin{align}\label{eq:Sdefq}
S^q_\bare(\pt)&=\frac{1}{N_c}\bra{0}\tr \left\{\bar{T}\left[S_{n}^\dagger S_{\bar{n}}\right]\delta^{(2)}(\pt-\mathcal{P}_{\perp}) \,T\left[S_{\bar{n}}^\dagger S_{n}\right]\right\}\ket{0},\\
S^g_\bare(\pt)&=\frac{1}{N_c^2-1}\bra{0}\tr \left\{\bar{T}\left[S_{n}^\dagger S_{\bar{n}}\right]\delta^{(2)}(\pt-\mathcal{P}_{\perp}) \,T\left[S_{\bar{n}}^\dagger S_{n}\right]\right\}\ket{0},
\label{eq:Sdefg}
\end{align}
where the $T$ and $\bar T$ are time and anti-time ordering operators and $\mathcal{P}$ is the SCET label momentum operator~\cite{Bauer:2001ct}.
The soft Wilson lines are exponentials of soft gauge fields. In momentum space we compactly write them as
\begin{align}
	S_n=\sum_{\mathrm{perm}}\exp\left[-\frac{g\, n\!\cdot\! A_s}{n \!\cdot\! \mathcal{P}}\right],
\end{align} 
where the sum is over all permutations of the momenta associated with the soft gluon modes~\cite{Bauer:2001yt}.
We leave it implicit in the notation of the Wilson lines that the gauge fields are in the fundamental representation of $SU(3)$ in the case of incoming quarks, or in the adjoint representation for incoming gluons. 
For later convenience, we have introduced an index $i\in\{q,g\}$ to denote the soft function and all the anomalous dimensions in the fundamental ($i=q$) or adjoint ($i=g$) representation, and the generic soft function $S^i_\bare(\pt)$.

Following \mycites{Stewart:2009yx,Stewart:2010qs}, the bare $n$-collinear beam functions are defined as spin-averaged forward proton matrix elements of $n$-collinear quark or gluon field operators ($z=\omega_-/\sqrt{s}$):
\begin{align}
\label{eq:Bq_def}
B_{q}^\bare(z,\mathbf{p}_{\perp})&=\theta(\omega_-)\bra{p_{n}}\bar{\chi}_{n}(0)\,\delta(\omega_- -\bn\!\cdot\! \mathcal{P}) \, \delta^{(2)}(\mathbf{p}_{\perp}-\mathcal{P}_{\perp})\frac{\slashed{n}}{2} \, \chi_n(0)\ket{p_n},\\
B_g^{\bare,\mu\nu}(z,\pt)&=\omega_- \, \theta(\omega_-)\bra{p_n}\tr \big\{\, \mathcal{B}^{\mu}_n(0) \,\delta(\omega_- - \bn\!\cdot\!\mathcal{P}) \, \delta^{(2)}(\pt-\mathcal{P}_{\perp})
\,\mathcal{B}_n^\nu(0) \big\}\ket{p_n},
\label{eq:Bg_def}
\end{align}
and analogous for the $\bn$ direction with $n \!\leftrightarrow\! \bn$, $\omega_- \!\!\leftrightarrow\! \omega_+$ as well as the antiquark beam function.
The $\omega_\pm$ is fixed by the delta function to the large light-cone momentum of the parton entering the hard interaction. 
We will often write $\omega_\pm$, when the discussion is equally valid for both directions. One should keep in mind though that different collinear directions have different large components and $\omega_+\omega_-=Q^2$. 
For more details on the relevant SCET definitions, as e.g.\ for the definitions of the (collinear) gauge invariant quark fields $\chi_n$ and gluon fields $\mathcal{B}_n^\mu$, we refer to \mycites{Bauer:2001yt,Chiu:2012ir,Stewart:2010qs}.

In order to separate non-perturbative contributions we write the beam functions as a convolution of perturbatively calculable matching coefficients $\mathcal{I}_{ij}$ and the usual collinear PDFs~\cite{Collins:1981uw,Fleming:2006cd,Stewart:2009yx}.
At leading order in $\lqcd^2/\pt^2$ we have for the renormalized TMD beam functions
\begin{align}\label{eq:Bq}
  B_q(z,\pt,\mu,\omega_\pm/\nu)&=\sum_{i}\mathcal{I}_{qi}(z,\pt,\mu,\omega_\pm/\nu)\otimes_z f_i(z,\mu),\\
  B_g^{\mu\nu}(z,\pt,\mu,\omega_\pm/\nu)&=\sum_{i}\bigg[\frac{g_\perp^{\mu\nu}}{2}\mathcal{I}_{gi}(z,\pt,\mu,\omega_\pm/\nu)
  \nn\\
  &+\left(\frac{g_\perp^{\mu\nu}}{2} + \frac{p_\perp^\mu p_\perp^\nu}{\pt^2}\right)\mathcal{J}_{gi}(z,\pt,\mu,\omega_\pm/\nu)\bigg]\otimes_z f_i(z,\mu)\,.
  \label{eq:Bg}
\end{align}
Here the sum is over all partons $i$, i.e.\ quarks and anti-quarks of all light flavors as well as gluons,
$z$ denotes the large light-cone momentum fraction w.r.t.\ the corresponding proton momentum ($P_\pm = \sqrt{s}$), i.e.\ $z=\omega_\pm/\sqrt{s}$, and $\mu$ is the UV renormalization scale.
We have introduced the symbol $\otimes_z$ for the (Mellin-) convolution
\begin{align}
  f(z)\otimes_z g(z)\equiv \int_z^1\frac{\df x}{x}f(x) \, g\Big(\frac{z}{x}\Big)\,.
\end{align} 
Our choice of rapidity regulator in addition introduces the rapidity renormalization scale $\nu$ as described below.

\subsection{Position space}
\label{subsec:bbspace}

In alternative to the formulation in momentum space, 
the relations and results can be formulated in position (impact parameter, $\bb$) space.
They are related by Fourier transformation in two dimensions:
\begin{align}
\label{eq:DefFTpttob}
\tilde{X}(\bb,\ldots)=\int \frac{\df^2 \pt}{(2\pi)^2} e^{i \bb\cdot\pt} X(\pt ,\ldots)
\,,
\\
X(\pt,\ldots)=\int \df^2 \bb \, e^{-i \bb\cdot\pt} \tilde{X}(\bb,\ldots)
\,,
\label{eq:DefFTbtopt}
\end{align}
where the ellipses denote any additional arguments.
Following a common abuse of notation, we will from now on drop the tilde and use the same symbols for the $\bb$ and $\pt$ space functions as the distinction is clear through their arguments.

\subsection{Rapidity regulator}
\label{subsec:regulator}

The beam and soft functions exhibit rapidity divergences that are not regulated by dimensional regularization. 
To regulate these divergences, 
we will use the rapidity regulator introduced in \mycites{Chiu:2012ir,Chiu:2011qc} in addition to dimensional regularization.
This regulator has been devised for covariant gauges and is defined by the following modification of the soft Wilson line,
\begin{align}
	S_n=\sum_{\mathrm{perm}}\exp\left[-\frac{g\, n\!\cdot\! A_s}{n \!\cdot\! \mathcal{P}} \,w\, \nu^{\eta/2} |2\mathcal{P}_{g3}|^{-\eta/2}\right],
\label{eq:RegSoftWilsonLine}
\end{align} 
and a related modification of the collinear Wilson line required for the regularization of the TMD beam functions.
In order to preserve gauge invariance and exponentiation, at higher loops we must not regulate the momentum of every gluon attached to the Wilson line individually.
Equation~\eqref{eq:RegSoftWilsonLine} instead should be interpreted to contribute a factor of $\propto w\, \nu^{\eta/2} |2\mathcal{P}_{g3}|^{-\eta/2}$ for every web attached to the Wilson line as a subdiagram. 
The (label) momentum operator $\mathcal{P}_{g3}=(\mathcal{P}_{g}^- - \mathcal{P}_{g}^+)/2$ then picks out the third component of the total (group) three-momentum flowing through the respective web.
Here, a web is understood as the maximally non-Abelian piece of a Feynman diagram, which cannot be disconnected by cutting two eikonal lines. After exponentiation, the exponent of the soft function is a sum of webs only.
For a precise definition of a web in the context of non-Abelian exponentiation we refer e.g.\ to \mycite{Berger:2002sv}. 

For our two-loop calculation of the TMD soft function the implementation in practice works as follows:
The parts of the diagrams, which are not fixed by the non-Abelian exponentiation theorem~\cite{Gatheral:1983cz,Frenkel:1984pz}, i.e.\ the ones with color factors other than $C_F^2$, we multiply with the regularization factor
\begin{align}\label{eq:regulator}
w^2\left(\frac{\nu}{|2\mathcal{P}_{g3}|}\right)^{\eta}\,.
\end{align}
Here we identify the group momentum $\mathcal{P}_{g}$ of the web with the total soft momentum 
crossing the final state cut.
In the $C_F^2$ pieces of the diagrams each gluon exchange individually represents a web and is regulated accordingly, see \subsec{exponentiation}.
Purely virtual diagrams vanish as scaleless integrals.

Our definition of the group momentum as the total momentum passing through the web across the cut is somewhat different to what is suggested in appendix A of \mycite{Chiu:2012ir}. There $\mathcal{P}_{g}$ is defined to give the web's total momentum transfer from the $S_{\bn}$ to the $S_n$ Wilson lines. The arguments given in \mycite{Chiu:2012ir} proving that gauge invariance and exponentiation are preserved by the $\eta$-regulator are however equally valid for our definition of the web's group momentum. 
Our choice allows us to recycle previously known results for the soft one-loop current in the calculation of the real-virtual diagrams in \subsubsec{single}.
As for the analytic regulator of \mycite{Becher:2011dz} with our definition gauge invariance is manifest, because effectively only the phase space integration is modified. Restricting $\mathcal{P}_{g}$ to act on real radiation only also allows for a sensible definition of a jet function.

The $\eta$-regulator works similarly to dimensional regularization: The bookkeeping parameter $w$ is the analogue of a renormalized coupling that will be set to one (for all values of $\nu$) in physical (rapidity finite) results, while $\nu$ is a renormalization scale similar to $\mu$ in dimensional regularization. The rapidity divergences manifest themselves as poles at $\eta=0$.

With this regulator, the soft and beam functions exhibit $1/\eta$ poles as well the ordinary $1/\e$ poles from dimensional regularization and it is crucial to send $\eta\rightarrow 0$ before $\e\rightarrow 0$ with $\eta/\e^n\rightarrow 0$ for all $n$ \cite{Chiu:2012ir}.
This is because at the bare level we first have to combine soft and beam functions such that the $\eta$ divergences cancel. We must then set $\eta\to 0$, before we can cancel their overall $1/\e$ poles with the ones from the 'bare' hard function, which is $\eta$ independent by construction.

Requiring the bare results to be $\nu$ independent leads to RRGEs for the renormalized beam and soft functions, that enable us to resum the rapidity logarithms~\cite{Chiu:2012ir}. 
For this to work obviously $w^2\nu^\eta$ in \eq{regulator} must be $\nu$-independent, and as a consequence
\begin{align}\label{eq:wBeta}
	\nu\frac{\df }{\df\nu}w=-\frac{\eta}{2} w\,,\qquad \lim_{\eta \to 0} w = 1\,.
\end{align}

The similarity to dimensional regularization also implies that with the $\eta$-regulator soft zero-bin contributions~\cite{Manohar:2006nz} in the (explicit) calculation of the beam functions amount to scaleless integrals and vanish.
This is also true for the analytic regulator proposed in~\mycite{Becher:2011dz} but does not hold e.g.\ for the delta regulator of~\mycite{Chiu:2009yx}.

Last but not least we would like to emphasize that besides \mycite{Stewart:2013faa}, our computation of the NNLO TMD soft function is the only two-loop calculation
and the first with an explicit demonstration of the non-Abelian exponentiation property, using the $\eta$-regulator.

It can therefore be considered a valuable consistency and practicability check of the rapidity regularization method beyond one loop.

\subsection{Transverse momentum in dimensional regularization}
\label{subsec:ptReg}

There are different schemes how to apply dimensional regularization when measuring the transverse momentum of the soft and collinear radiation, see e.g.\ the discussion in \mycite{Jain:2011iu}.
For the calculation of the soft function as presented in this paper, we will adopt the CDR$_2$ scheme, where the transverse momentum, $\pt$, of the soft and beam functions is a two-dimensional vector, while all transverse loop momenta are $(d-2)$-dimensional vectors.\footnote{Here and in the following we refer also to the momenta of the soft partons that cross the final state cut as loop momenta.} 
The delta functions to measure the transverse momentum in \eqsc{Sdefq}{Sdefg} and 
\eqsc{Bq_def}{Bg_def} 
therefore only fix two components of the ($d-2$)-dimensional transverse loop momenta in the perturbative evaluation of the soft and beam function operator matrix elements.
This scheme ensures that the soft and beam functions, and therefore the factorization formulas for the $\pt$ distribution in \eqs{factq}{factg}, live in integer dimensions.

Due to the isotropic nature of the soft radiation, the soft function $S(\pt)$ is independent of the orientation of the vector $\pt$ in the transverse plane.\footnote{An analogous argument holds for the quark beam function, but not for the gluon beam function, \eq{Bg}, which has a nontrivial Lorentz (tensor) structure.}
We can therefore replace the delta function in \eqs{Sdefq}{Sdefg} as
\begin{align}
\label{eq:deltaReplace}
	\delta^{(2)}(\pt-\mathcal{P}_\perp) \rightarrow \frac{1}{\pi}\delta(\pt^2-\mathcal{P}_\perp^2)\,,
\end{align}
where the factor $1/\pi$ on the right hand side (RHS) compensates for the contribution from the azimuthal integration, when taking the implicit sum over all soft final states, whose total transverse momentum is collected by the (label) momentum operator $\mathcal{P}_\perp$.

The replacement in \eq{deltaReplace} gives rise to a subtlety concerning the dimensional regularization of the soft function. 
The measurement functions on both sides of \eq{deltaReplace} only yield identical results for the bare soft function in $d$ dimensions
if the momentum operator $\mathcal{P}_\perp$ on the RHS is interpreted to return only two transverse momentum components, while the residual $(d-4)$ components of the total soft transverse momentum crossing the final state cut remain unrestricted.

The definition of the soft function using the RHS of \eq{deltaReplace} however also admits a different scheme for the dimensional regularization.
In particular we can interpret $\mathcal{P}_\perp^2$ as the square of the full $(d-2)$ dimensional transverse momentum of the soft radiation.
In this case the scalar delta function corresponds to the vector-like measurement
\begin{align}
\frac{(\pt^2)^{-\epsilon}}{\Gamma(1-\epsilon)\pi^{\epsilon}}\; \delta^{(d-2)}(\pt-\mathcal{P}_\perp) \,,
\label{eq:OtherScheme}
\end{align}
where also $\pt$ is promoted to $(d-2)$ dimensions in the definition of the bare soft functions.
The latter will now however differ from the ones defined in CDR$_2$, i.e.\ \eqs{Sdefq}{Sdefg}.

Both schemes are nevertheless equally viable for the calculation of the soft function.
To see this, we note that the $1/\epsilon^n$ poles in the bare soft function originate from virtual ultraviolet (UV) divergences, which evade the cancellation by UV divergent real corrections because of the $\pt$ constraint.
As a consequence, all NLO $1/\epsilon^n$ divergences ($n=1,2$) are proportional to $\delta(\pt^2)$, cf.\ \eq{ZNLO}.
Since the divergent one-loop virtual corrections (where $\mathcal{P}_\perp \equiv 0$) are not affected by the measurement function we find the same bare result up to $\ord(\eps)$ terms in both schemes discussed above.

At higher orders, starting from NNLO, there are mixed real-virtual contributions. The bare expressions for the $\pt$-dependent soft function will therefore in general depend on the scheme. We indeed find different subleading terms in $1/\eps$.
Upon renormalization these differences however cancel out and we obtain scheme independent results for the renormalized soft functions and their anomalous dimensions.
We have explicitly verified this by performing the complete calculation of the two-loop soft function in both schemes, CDR$_2$ and using the measurement according to \eq{OtherScheme}.
This also serves as a cross check of our final NNLO results.

\section{Renormalization and logarithmic structure}
\label{sec:logstruc}

In this section we briefly review the (rapidity) renormalization of the beam and soft functions according to \mycite{Chiu:2012ir} and predict their logarithmic structure based on the (R)RGEs.

The renormalized soft and beam functions and their respective renormalization factors, $Z^i_{S,B}$, that absorb all UV and rapidity divergences are defined through
\begin{align}\label{eq:Sbare2}
S_\bare^i(\pt) &=Z_S^i(\pt,\mu,\nu)\otimes_\perp S^i(\pt,\mu,\nu)\,, \\
B^\bare_i(z,\pt) &=  Z_B^i(\pt,\mu,\omega_\pm/\nu) \otimes_\perp B_i(z,\pt,\mu,\omega_\pm/\nu)\,,
\end{align}
where we have introduced the generic notation with the index $i\in\{q,g\}$ also for the beam function. Lorenz indices for the gluon beam function ($i=g$) are understood.
The symbol $\otimes_\perp$ denotes the convolution
\begin{align}
\label{eq:ptConv}
g(\pt)\otimes_\perp f(\pt)\equiv \int\!\!\frac{\df^2q_\perp}{(2\pi)^2} \,f(\mathbf{p}_\perp-\mathbf{q}_\perp) \, g(\mathbf{q}_\perp)
\end{align}
in two-dimensional transverse momentum space.

From the above renormalization factor $Z^i_{S}$ we can derive the anomalous dimensions for the soft RGE and RRGE in momentum space,
\begin{align}
\label{eq:gamma_mu_def}
\gamma_{S,\mu}^i(\pt,\mu,\nu)&=-\Big[Z_S^i(\pt,\mu,\nu)\Big]^{-1}\otimes_\perp\Big[\mu\frac{\df}{\df\mu}Z_S^i(\mathbf{p}_{\perp},\mu,\nu)\Big]\,,\\
\gamma_{S,\nu}^i(\pt,\mu) &= 
-\Big[Z_S^i(\pt,\mu,\nu)\Big]^{-1} \otimes_\perp \Big[\nu\frac{\df}{\df\nu}Z_S^i(\mathbf{p}_{\perp},\mu,\nu)\Big]\,,
\label{eq:gamma_nu_def}
\end{align}
and analogous for the beam function anomalous dimensions.
The $\mu$ anomalous dimensions actually only have a trivial $\pt$ dependence \cite{Chiu:2012ir} as elaborated on below \eq{munurelation}.
To make this explicit,
we can write
\begin{align}
\label{eq:pTgamma_Smu}
\gamma_{S,\mu}^i(\pt,\mu,\nu) = (2\pi)^2\ptdelta\gamma_{S,\mu}^i(\mu,\nu)
=
(2\pi)^2\ptdelta\sum_{n=0}^\infty \gamma_{S,\mu}^{i(n)}(\mu,\nu)\,a_s^{n+1}
\,,
\end{align}
and analogous for $\gamma_{B,\mu}^i$.
For convenience when expressing functions as power series in $\alpha_s$, we define
\begin{align}
 a_s \equiv \asmu.
\end{align}
We expand the renormalized soft function and its renormalization factor as
\begin{align}
\label{eq:S_coef}
  S^i(\pt,\mu,\nu)=\sum_{n=0}^\infty {S^i}^{(n)}(\mathbf{p}_{\perp},\mu,\nu) \, a_s^n,	&&  Z_S^i(\pt,\mu,\nu)=\sum_{n=0}^\infty {Z_S^i}^{(n)}(\mathbf{p}_{\perp},\mu,\nu) \, a_s^n \,.
\end{align}
The LO (tree-level) coefficients are
\begin{align}
	{S^q}^{(0)}(\pt)={S^g}^{(0)}(\pt)=\delta^\two(\pt)\,,	&&	{Z_S^q}^{(0)}(\pt)={Z_S^g}^{(0)}(\pt)=(2\pi)^2 \delta^\two(\pt)\,.
\end{align}

\subsection{Recurrence relations}
\label{subsec:logStructure}

The (R)RG structure of the soft and beam functions is conveniently discussed in impact parameter ($\bb$) space. 
Upon the Fourier transformation in \eq{DefFTpttob}, convolutions of the type in \eq{ptConv} turn into ordinary products.
For notational convenience we define
\begin{align}
\label{eq:DefLb}
   L_b\equiv\ln\left(\frac{\mathbf{b}^2\mu^2e^{2\gamma_E}}{4}\right) .
\end{align}
All $\bb$ space expressions given below can be straightforwardly transformed back to $\pt$ space.%
\footnote{Given $\mu_{(B,S)}$ are chosen independent of $\bb$.}
The two-dimensional Fourier transform of powers of the logarithm in \eq{DefLb} can be expressed in terms of plus distributions $\plus{n}(\pt,\mu)$ defined in \app{plusDistributions}, where we also give corresponding translation tables.

The RGE and RRGE of the soft and beam functions in $\bb$ space read
\begin{align}
\mu\frac{\df}{\df\mu}S^i(\bb,\mu,\nu)&={\gamma_{S,\mu}^i}(\mu,\nu) \, S^i(\bb,\mu,\nu)\,, \label{eq:softRGE}\\
\nu\frac{\df}{\df\nu}S^i(\bb,\mu,\nu)&={\gamma_{S,\nu}^i}(\bb,\mu) \, S^i(\bb,\mu,\nu)\,, \label{eq:softRRGE} \\
\mu\frac{\df}{\df\mu}B_i(z,\bb,\mu,\omega_\pm/\nu)&={\gamma_{B,\mu}^i}(\mu,\omega_\pm/\nu) \, B_i(z,\bb,\mu,\omega_\pm/\nu)\,, \label{eq:beamRGE}\\
\nu\frac{\df}{\df\nu}B_i(z,\bb,\mu,\omega_\pm/\nu)&=\gamma_{B,\nu}^i(\bb,\mu) \, B_i(z,\bb,\mu,\omega_\pm/\nu)\,.
\label{eq:beamRRGE}
\end{align}
The FO cross section must be independent of $\mu$ and $\nu$ order by order in $\alpha_s$.
The anomalous dimensions must therefore fulfill the consistency relations
\begin{align}
\label{eq:murelation}
\gamma_{H,\mu}^i(Q,\mu)+\gamma_{S,\mu}^i(\mu,\nu)+\gamma_{B,\mu}^i(\mu,\omega_+/\nu)+\gamma_{B,\mu}^i(\mu,\omega_-/\nu)&=0\,,\\
\label{eq:munurelation}
\gamma_{S,\nu}^i(\bb,\mu)+2\gamma_{B,\nu}^i(\bb,\mu)&=0\,,
\end{align}
where $\gamma_{H,\mu}^i$ denotes the anomalous dimension of the hard function. Besides $\mu$ the latter only depends on the hard scale $Q$.
In \eqs{softRGE}{beamRGE} we therefore anticipated that, due to \eq{murelation} and the fact that in SCET$_{\rm II}$ the soft and collinear sectors are located on the same invariant mass hyperbola, the $\mu$ anomalous dimensions ${\gamma_{S,\mu}^i}$ and ${\gamma_{B,\mu}^i}$ are also independent of $\bb$~\cite{Chiu:2012ir}.
Given the well-known Sudakov form of $\gamma_{H,\mu}^i$ we can thus infer the structure of ${\gamma_{S,\mu}^i}$ and ${\gamma_{B,\mu}^i}$ from \eq{murelation},
\begin{align}\label{eq:gamma_S_i}
{\gamma_{S,\mu}^i}(\mu,\nu)&=4\Gamma^i_\cusp\ln\frac{\mu}{\nu}+\gamma_S^i
\,,\\
{\gamma_{B,\mu}^i}(\mu,\omega_\pm/\nu)&=2\Gamma^i_\cusp\ln\frac{\nu}{\omega_\pm}+\gamma_B^i
\label{eq:gamma_B_i}\,.
\end{align}
The universal cusp and the soft and beam non-cusp anomalous dimensions
\begin{align}
\label{eq:ExpGams}
 \Gamma^i_\cusp = \sum_{n=0}^\infty \Gamma^i_n \, a_s^{n+1} \,,
&&
\gamma^i_S = \sum_{n=0}^\infty {\gamma_S^i}_n \, a_s^{n+1} \,,
&&
\gamma^i_B = \sum_{n=0}^\infty {\gamma_B^i}_n \, a_s^{n+1} \,,
\end{align}
only depend on $\mu$ through $a_s$.
The coefficients $\Gamma^i_{n}$ are known up to three loops \cite{Korchemsky:1987wg,Moch:2004pa}. We write $\Gamma^q_n/C_F = \Gamma^g_n/C_A = \Gamma_n$ with 
\begin{align}
\label{eq:gamma0}
\Gamma_0&=4\,,& 
\Gamma_1&=\bigg(\frac{268}{9}-\frac{4\pi^2}{3}\bigg)C_A-\frac{80}{9}T_Fn_f\,.
\end{align}
From \eqsc{munurelation}{gamma_S_i} and because derivatives commute, we have
\begin{align}
\mu\frac{\df}{\df\mu}{\gamma_{S,\nu}^i}=\nu\frac{\df}{\df\nu}{\gamma_{S,\mu}^i}=-2\,\nu\frac{\df}{\df\nu}{\gamma_{B,\mu}^i}&=-4\Gamma_{\mathrm{cusp}}^i\,.
\label{eq:rel2}
\end{align}
Given the simple relation in \eq{munurelation}, we define a common $\nu$ anomalous dimension 
\begin{align}
\gamma_\nu^i(\bb,\mu) 
= \sum_{n=0}^\infty{\gamma_\nu^i}^{(n)}(\bb,\mu)\,a_s^{n+1}
\equiv \gamma_{S,\nu}^i(\bb,\mu) = -2{\gamma_{B,\nu}^i} (\bb,\mu)\,.
\end{align}
Equation \eqref{eq:rel2} allows us to determine the structure of $\gamma_\nu^i$.
Expanding it we find for the $n$-th coefficient
\begin{align}
\label{eq:rel_gamma_nu_n}
\mu\frac{\df}{\df\mu}{\gamma_\nu^i}^{(n)}(\bb,\mu)&=-4\Gamma_n^i +2\sum_{m=0}^{n-1}(m+1)\,\beta_{n-m-1}\, {\gamma_\nu^i}^{(m)}(\bb,\mu)\,,
\end{align}
where the $\beta_i$ denote the usual QCD $\beta$-function coefficients with $\beta_0=(11C_A-4 T_F n_f)/3$. 
Upon integration we thus obtain the recurrence relation
\begin{align}\label{eq:nurecurrence}
	{\gamma_\nu^i}^{(n)}(\bb,\mu)&=-2\Gamma_n^i\ln\frac{\mu^2}{\mu_S^2}+2\sum_{m=0}^{n-1}(m+1)\beta_{n-m-1}\int_{\mu_S}^\mu \frac{\df\mu'}{\mu'} {\gamma_\nu^i}^{(m)}(\bb,\mu')+{\gamma_\nu^i}_n\,,
\end{align}
with ${\gamma_\nu^i}_n \equiv {\gamma_\nu^i}^{(n)}(\bb,\mu_S)$.

Analogous to this derivation, the RGE and RRGE for the soft function in \eqs{softRGE}{softRRGE} imply differential equations for its expansion coefficients.
Integrating them, we arrive at%
\footnote{This for our purpose convenient form has been suggested by Markus Ebert in private communication.}
\begin{align}\label{eq:Srecurrence}
{S^i}^{(n)}(\bb,\mu,\nu)&=
\sum_{m=0}^{n-1}\Big\{\left[2m\beta_{n-m-1}+{\gamma_S^i}_{n-m-1}\right]\int_{\mu_S}^\mu\frac{\df\mu'}{\mu'} \, {S^i}^{(m)}(\bb,\mu',\nu) \nn\\
&
+4\Gamma_{n-m-1}^i\int_{\mu_S}^\mu\frac{\df\mu'}{\mu'}\ln\frac{\mu'}{\nu} \; {S^i}^{(m)}(\bb,\mu',\nu)\nn\\
&
+\,{\gamma_\nu^i}_{n-m-1}
\int_{\nu_S}^{\nu}\frac{\df\nu'}{\nu'} \, {S^i}^{(m)}(\bb,\mu_S,\nu')\Big\}+S^i_n \,,
\end{align}
for the expansion coefficients of the soft function according to the Fourier transformed version of \eq{S_coef}.
The results in \eqs{nurecurrence}{Srecurrence} involve the integration constants
\begin{align}
\label{eq:Sintconstants}
S^i_n \equiv {S^i}^{(n)}(\bb,\mu_S,\nu_S)\,, && {\gamma_\nu^i}_n \equiv {\gamma_\nu^i}^{(n)}(\bb,\mu_S) \,.
\end{align}
We can choose the scales $\mu_S$ and $\nu_S$ such that 
these integration constants
are free of large logarithms and even become $\bb$-independent numbers e.g.\ by setting them to their canonical values, given below in \eq{defaultscales}.
Also note that for the canonical value of $\mu_S$ we have $\ln\mu^2/\mu_S^2 = L_b$.
In this way all RG and RRG logarithms ($\propto L_b^n$) are generated by the recursion.
We can therefore use \eqs{nurecurrence}{Srecurrence} to predict the logarithmic structure of the $\nu$ anomalous dimension and the soft function, respectively. 
Through NNLO we find perfect agreement with the results of our explicit two-loop calculation of the soft function in~\subsec{bspace}.

The structure of the beam function matching coefficients $\mathcal{I}_{ij}$ and $\mathcal{J}_{gj}$ in \eqs{Bq}{Bg} can analogously be derived from the RGE and RRGE for the beam functions, \eqs{beamRGE}{beamRRGE}.
To do so, we first note that the PDFs obey the DGLAP equations
\begin{align}
\label{eq:DGLAP}
	\mu\frac{\df}{\df\mu}f_i(z,\mu)=2\sum_jP_{ij}(z,\mu)\otimes_zf_j(z,\mu)\,,
\end{align}
where the splitting functions can be expressed as power series in $a_s$,
\begin{align}
	P_{ij}(z,\mu)=\sum_{n=0}^\infty(2a_s)^{n+1}P_{ij}^{(n)}(z)\,.
\end{align}
All relevant $P_{ij}^{(n)}(z)$ functions in the notation used here as well as convolutions among them, can e.g.\ be found in the appendices of \mycites{Gaunt:2014xga,Gaunt:2014cfa}. 
From \eqsc{beamRGE}{beamRRGE} and \eq{DGLAP} we obtain the following RGE and RRGE for the matching coefficients:
\begin{align}\label{eq:IRGE}
  \mu\frac{\df}{\df\mu}\mathcal{I}_{ij}(z,\bb,\mu,\omega_\pm/\nu)&
  =\sum_k\mathcal{I}_{ik}(z,\bb,\mu,\omega_\pm/\nu)
  \otimes_z\left[\delta(1-z)\delta_{kj} \, {\gamma_{B,\mu}^i}(\mu,\omega_\pm/\nu)-2P_{kj}(z,\mu)\right],\\\label{eq:IRRGE}
  \nu\frac{\df}{\df\nu}\mathcal{I}_{ij}(z,\bb,\mu,\omega_\pm/\nu)&
  =-\frac{1}{2}\gamma^i_\nu(\bb,\mu) \, \mathcal{I}_{ij}(z,\bb,\mu,\omega_\pm/\nu)\,.
\end{align}
We expand
\begin{align}
\label{eq:IijbspaceExp}
  \mathcal{I}_{ij}(z,\bb,\mu,\omega_\pm/\nu)&=\frac{1}{(2\pi)^2}\sum_{n=0}^\infty \mathcal{I}_{ij}^{(n)}(z,\bb,\mu,\omega_\pm/\nu)\,a_s^n\,.
\end{align}
Just as for $S^i$, we obtain a recurrence relation for the coefficients $\mathcal{I}_{ij}^{(n)}$ by integrating 
the differential equations for these coefficients following from
\eqs{IRGE}{IRRGE}:
\begin{align}\label{eq:Irecurrence}
  \mathcal{I}_{ij}^{(n)}(z,\bb,\mu,\omega_\pm/\nu)&=\sum_{m=0}^{n-1}\Bigg[
  \sum_\ell
  \int_{\mu_B}^\mu\frac{\df\mu'}{\mu'}\mathcal{I}_{il}^{(m)}(z,\bb,\mu',\omega_\pm/\nu) 
  \otimes_z
  \bigg\{
  - 2^{n-m+1}P_{\ell j}^{(n-m-1)}(z) \nn\\
  &
  +\delta(1-z)\delta_{lj}\left(2\Gamma_{n-m-1}^i\ln\frac{\nu}{\omega_\pm}+{\gamma_B^i}_{n-m-1}+2m\beta_{n-m-1}\right) \bigg\} \nn\\
  &
  -\frac{1}{2}{\gamma_\nu^i}_{n-m-1} 
  \int_{\nu_B}^\nu\frac{\df\nu'}{\nu'}\mathcal{I}_{ij}^{(m)}(z,\bb,\mu_B,\omega_\pm/\nu')\Bigg]
  +I_{ij}^{(n)}(z)\,,
\end{align}
with the integration constant
\begin{align}
\label{eq:Iijconst}
I_{ij}^{(n)}(z) \equiv \mathcal{I}_{ij}^{(n)}(z,\bb,\mu_B,\omega_\pm/\nu_B^\pm)\,.
\end{align}
Equations~\eqref{eq:IRGE}-\eqref{eq:Irecurrence} hold in full analogy also for $\mathcal{J}_{gj}$.
Again, setting $\mu_B$ and $\nu_B$ to their canonical values in \eq{defaultscales} the $I_{ij}^{(n)}(z)$ and $J_{gj}^{(n)}(z)$
become $\bb$ independent and are pure functions of $z$.
We will use the recursion relations to derive the logarithmic structure of all beam function matching kernels through NNLO.
Note that to this order all ingredients of \eq{Irecurrence} but the $I^{(n)}_{ij}(z)$ are known, once we have determined the soft (non-cusp) and $\nu$ anomalous dimensions from the direct calculation in \sec{calculation}.

By matching our expressions onto the beam functions calculated in another (rapidity regularization) scheme, we are able to fix the missing coefficients $I_{ij}^{(n)}(z)$ and $J_{gj}^{(n)}(z)$ and hence obtain the full NNLO beam functions in the RRG scheme, see~\sec{beam}.\footnote{Actually we are matching cross sections involving also the soft function. For details see \subsec{rel_BN}.}
In fact the predicted structure of the beam functions provides a strong cross check of the extracted result.

\subsection{Resummation}
\label{subsec:resummation}
In addition to using the RGE and RRGE of $S^i$ and $B_i$ to obtain the logarithmic structure at each FO in $\alpha_s$,
we can also use them to resum the logarithms to all orders in $\alpha_s$.
The RGE and RRGE for $X=S^i,B_i$ both have the form 
\begin{align}
 s \frac{\df}{\df s} \ln X(s) = \gamma_{X, s}(s) \,,
\end{align}
with the generic solution
\begin{align}
\label{eq:resum_formal_sol}
X(s) = X(s_0)\exp\left[ \int_{s_0}^{s} \frac{\df s'}{s} \gamma_{X, s}(s') \right] .
\end{align}
Since $\gamma_{S,\nu}^i\,(=-\tfrac{1}{2}\gamma_{B, \nu}^i)$ is independent of $\nu$,
the solutions of the RRGEs simplify to
\begin{align}
 \label{eq:S_logNu_resum}
  S^i(\bb,\mu,\nu)&=\exp\left(\gamma_\nu^i(\bb,\mu) \, \ln\frac{\nu}{\nu_S}\right)S^i(\bb,\mu,\nu_S)\,,\\
  B_i(z,\bb,\mu,\omega_\pm/\nu)&=\exp\left(-\frac{1}{2}\gamma_\nu^i(\bb,\mu)\, \ln\frac{\nu}{\nu_B^\pm}\right)B_i(z,\bb,\mu,\omega_\pm/\nu_B^\pm)\,,
 \label{eq:B_logNu_resum}
\end{align}
where only a single logarithm of $\nu$ appears in the exponent of the resummation factors.
Accordingly, the maximum power of the rapidity logarithms at $\ord(\alpha_s^n)$ in the cross section is $n$.
This is in contrast to the series of double (Sudakov) logarithms generated by the $\mu$ anomalous dimensions which themselves depend on $\mu$ and originate from the interplay of soft and collinear singularities.
Since the ordinary PDF $f_i$ is independent of $\nu$, the analogue of \eq{B_logNu_resum} holds for the matching kernels $\mathcal{I}_{ij}$ in \eqs{Bq}{Bg}.

Choosing $s_0$ in \eqref{eq:resum_formal_sol} of order of the corresponding 'natural' scale, allows for the resummation of large logarithms.
The natural scales are given by the requirement that the logarithms in the FO expressions $S^i(\bb,\mu_S,\nu_S)$ and $B_i(z,\bb,\mu_B,\omega_\pm/\nu_B)$ are rendered small.
The canonical choice in $\bb$ space is
\begin{align}\label{eq:defaultscales}
\mu_B = \mu_S = \nu_S = 2 \frac{e^{-\gamma_E}}{|\bb|}\,,&&\nu_B^\pm = \omega_\pm
\,.
\end{align}
Using these scales, all (R)RG logarithms in the FO results for beam and soft functions, i.e.\ $L_b$, $\ln (\mu_S/\nu_S)$ and $\ln (\nu_B/\omega_\pm)$, vanish, cf.\ \eqsc{Sbspace1}{Sbspace2} and \eqss{I1}{J2}.

\section{Calculation of the soft function}
\label{sec:calculation}
In this section we compute the soft function in \eq{Sdefq} to NNLO.
We perform the calculation in momentum space and with the Wilson lines in the fundamental color representation.
Our results are straightforwardly generalized to an arbitrary color representation due to Casimir scaling and translated to impact parameter space. 
At two loops, we separately treat the contributions with non-$C_F^2$ (i.e.\ $C_A C_F$ and $C_F T_F n_f$) and $C_F^2$ color factor. For the latter we explicitly verify non-Abelian exponentiation in the presence of the $\eta$-regulator, see \subsec{exponentiation}. 
Throughout this paper, we use the \MSbar (UV) renormalization scheme and Feynman gauge for the calculations. 
The \MSbar relation between bare and renormalized couplings is
\begin{align}\label{eq:aRen}
\frac{g_b^2}{4\pi}=\frac{g^2 Z_g^2 }{4\pi}\left(\frac{\mu^2 e^{\gamma_E}}{4\pi}\right)^{\e} = \mu^{2\e}a_s(4\pi)^{1-\e}e^{\e \gamma_E}\Big(1-\beta_0\, a_s\frac{1}{\e}+\mathcal{O}(a_s^2)\Big).
\end{align}
The bare results presented in this section are obtained for the measurement \eq{deltaReplace} in the CDR$_2$ scheme.
The renormalized results are independent of this scheme, see discussion in \subsec{ptReg}.
An extended discussion of several aspects is given in \mycite{Joelthesis}.

\subsection{NLO}
\label{subsec:nlo}
\begin{figure}[t]
\begin{center}
\begin{tabular}{cc}
\includegraphics[width=4cm]{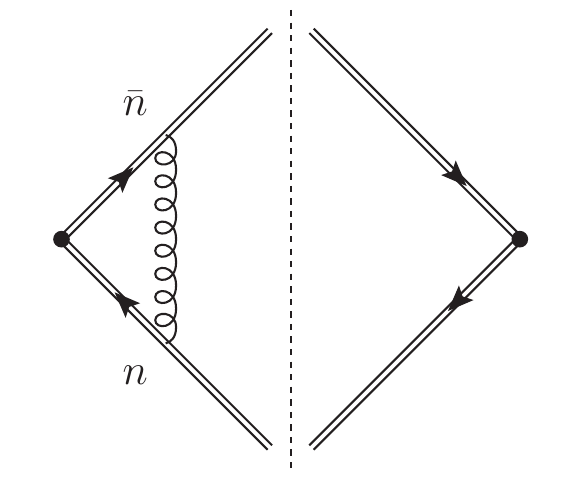}&
\includegraphics[width=4cm]{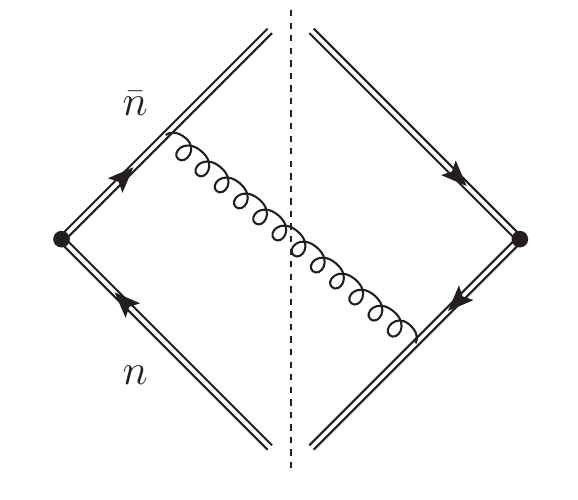}\\
(a)	&	(b)
\end{tabular}
\caption{Soft one-loop diagrams. 
The double lines represent Wilson lines in the fundamental representation. The little arrows indicate the fermion flow of the corresponding full theory quarks. The dotted line represents the final state cut.
Every cut propagator corresponds to a real emission. Diagram (a) is scaleless and vanishes in dimensional regularization. 
Diagrams with gluons attached only to Wilson lines in the same light-like direction vanish as $n^2=\bn^2=0$.
Diagram (b) and its mirror graph are therefore the only diagrams contributing to the soft function at NLO. 
\label{fig:nlo}}
\end{center}
\end{figure}

The NLO soft function in the RRG formalism has been calculated in the adjoint representation in \mycite{Chiu:2012ir}.
The relevant one-loop diagrams are shown in \fig{nlo}. In agreement with \mycite{Chiu:2012ir} we obtain
\begin{align}
& S^q_{\bare,\rm 1loop}(\pt)=2(n\cdot\bar{n})g_b^2w^2\nu^\eta\frac{\mathrm{Tr}(T^aT^a)}{N_c}\int\frac{\df^dk}{(2\pi)^d}|k^--k^+|^{-\eta}\frac{(2\pi)\delta^+(k^2)\delta^{(2)}(\mathbf{p}_{\perp}-\mathbf{k}_\perp)}{k^-k^+}\nn\\
&=\left[a_s- a_s^2\frac{\beta_0}{\e}+\mathcal{O}(a_s^3)\right]
 \frac{4 C_Fw^2\nu^\eta\mu^{2\epsilon}e^{\e\gamma_E}}{\pi}\frac{\Gamma\left(1+\e+\frac{\eta}{2}\right)}{\Gamma\left(1+\frac{\eta}{2}\right)}\frac{\Gamma(\frac{1}{2}-\frac{\eta}{2})\Gamma(\frac{\eta}{2})}{2^\eta\sqrt{\pi}} (\pt^2)^{-1-\epsilon-\eta/2}
 \label{eq:nlobare}
\end{align}
for the bare result in the fundamental representation.
In the second step we performed the integration and expressed $g_b$ through the renormalized coupling $a_s$ using \eq{aRen}. This directly reveals the NNLO contribution from the one-loop calculation, which is proportional to $\beta_0$.
In \subsec{exponentiation} we will show that the square of \eq{nlobare} (in a convolutional sense) determines the $C_F^2$ part of the NNLO result. 

\subsection{NNLO}
\label{subsec:nnlo}

The two-loop diagrams contributing at NNLO can be divided into two categories: real-virtual (single real emission) and real-real (double real emission) graphs. They are displayed in \figs{rv}{rr}, respectively, and their color structure can be read off \Tab{color}. 
As previously stated, we ignore purely virtual diagrams since they vanish for our choice of regulators. 
\begin{table}[h!]
\centering
\begin{tabular}{|c|c|}
  \hline
  Color Factor 	&	Diagrams	\\
  \hline
  $C_F$	& 1-loop\\
  $C_F^2$	&	$\mathcal{I}$, $\mathcal{R}$\\
  $C_FC_A$	&	$\mathcal{I}$, $\mathcal{T}$, $\mathcal{G}$, $\mathcal{H}$, $\mathcal{R}$\\
  $C_FT_Fn_f$	&	$\mathcal{Q}$ \\
  \hline
\end{tabular}
\caption{Table of diagrams (2nd column) and the color factors they involve (1st column).
The corresponding diagrams are displayed in \fig{nlo}, \ref{fig:rv} and \ref{fig:rr}. 
After renormalization, the one-loop diagram contributes to both $C_FC_A$ and $C_F T_F n_f$ through the $\beta_0$ term in \eq{aRen}.}
\label{tab:color}
\end{table}

\subsubsection{Single real emission}
\label{subsubsec:single}

\begin{figure}[t]
\begin{center}
\begin{tabular}{ccc}
\includegraphics[width=4cm]{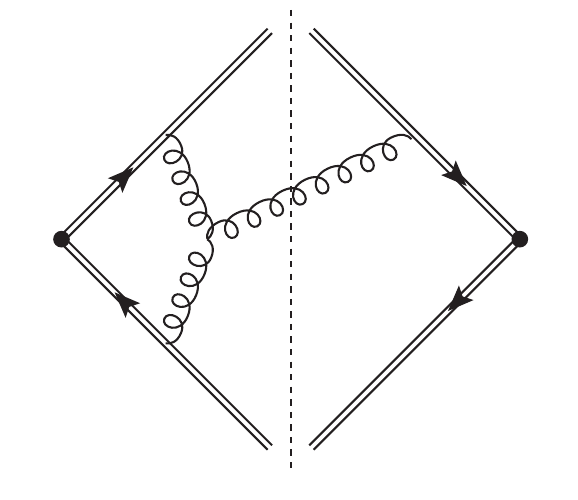}&
\includegraphics[width=4cm]{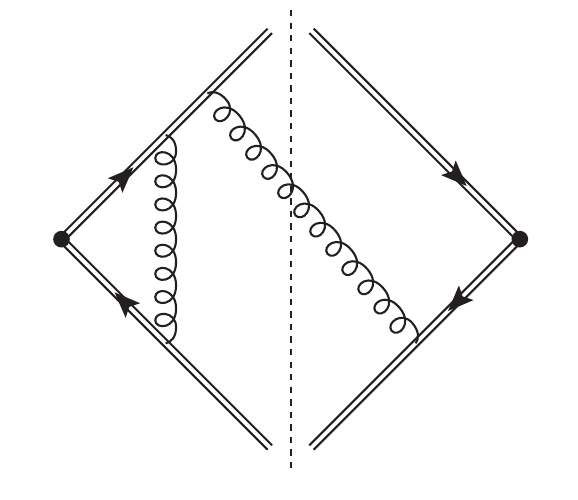}&
\includegraphics[width=4cm]{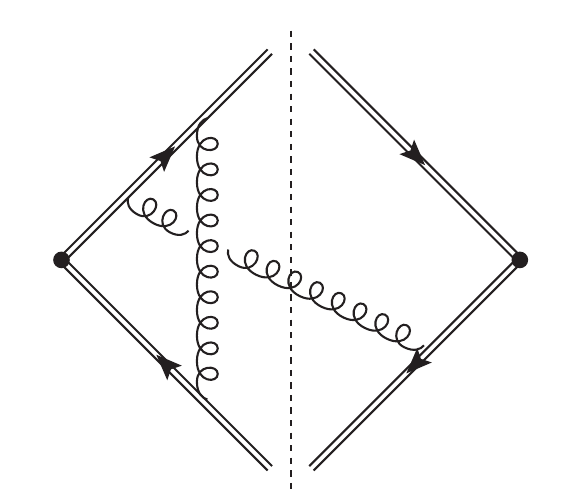}\\
(a)	&	(b)	&	(c)
\end{tabular}
\caption{Soft single real emission (real-virtual) diagrams contributing at NNLO. 
Their sum, including all mirror graphs, is denoted by $\mathcal{R}$.
The $C_F^2$ piece of $\mathcal{R}$, from diagrams (b) and (c), vanishes for our choice of regulators. The $C_FC_A$ piece, from diagrams (a) and (c), we calculate using the known one-loop soft current.}
\label{fig:rv}
\end{center}
\end{figure}

We use the known result for the one-loop soft current~\cite{Catani:2000pi} to calculate the $C_FC_A$ part of the sum of real-virtual diagrams ($\mathcal{R}$) in \fig{rv}. 
In the end, the calculation differs from the one-loop calculation in \eq{nlobare} only by an overall factor in the integrand:
\begin{align}
\mathcal{R}_{C_FC_A}&=-\frac{a_s^2 C_F C_A w^2 2^{7+2 \epsilon} \pi^{2+\epsilon } \nu^{\eta } \mu^{2 \epsilon } \Gamma^4(1-\epsilon ) \Gamma^3(1+\epsilon )}{\epsilon^2 \Gamma^2(1-2 \epsilon ) \Gamma(1+2 \epsilon)}\left(\frac{e^{\gamma_E}}{4\pi}\right)^{2\e}\nonumber\\
   &\times\int\frac{\df^dk}{(2\pi)^d}|k^--k^+|^{-\eta}\frac{(2\pi)\delta^+(k^2)\delta^{(2)}(\mathbf{p}_{\perp}-\mathbf{k}_\perp)}{(k^-k^+)^{1+\epsilon}}\nn\\
&=-a_s^2 C_F C_A 2^{3 - \eta} \pi^{-3/2} e^{2 \e \gamma_E} w^2 \mu^{4 \e} \nu^\eta\nn\\
&\times\frac{\Gamma(\frac{1}{2}-\frac{\eta}{2})\Gamma(\frac{\eta}{2})\Gamma^4(1-\epsilon) \Gamma^3(1+\epsilon)}{\epsilon^2\Gamma^2(1-2\epsilon) \Gamma(1+2\epsilon)}\frac{ \Gamma \left(1+2 \epsilon +\frac{\eta }{2}\right)}{\Gamma \left(1+\epsilon +\frac{\eta }{2}\right)}\, (\pt^2)^{-1-2\epsilon-\frac{\eta}{2}}
\,.
\label{eq:rCFCAbare}
\end{align}

\subsubsection{Double real emissions}
\label{subsubsec:double}

\begin{figure}[h!]
\begin{center}
\begin{tabular}{ccc}
\includegraphics[width=4cm]{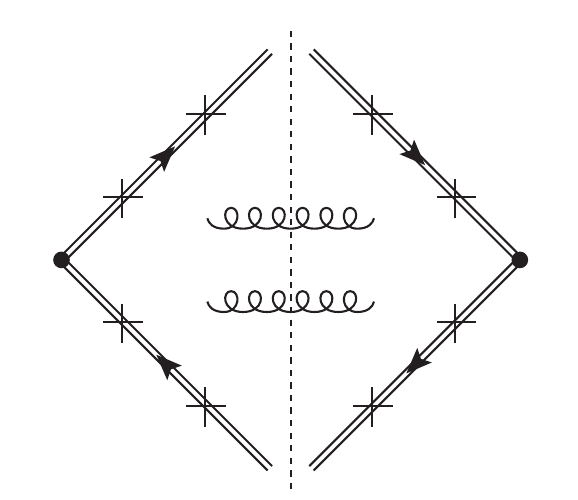}&
\includegraphics[width=4cm]{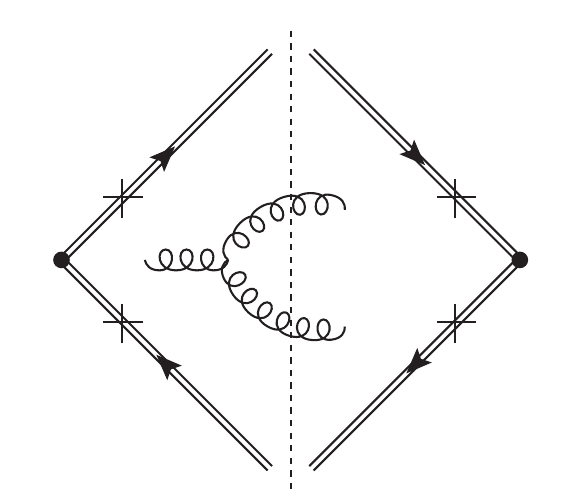}&
\includegraphics[width=4cm]{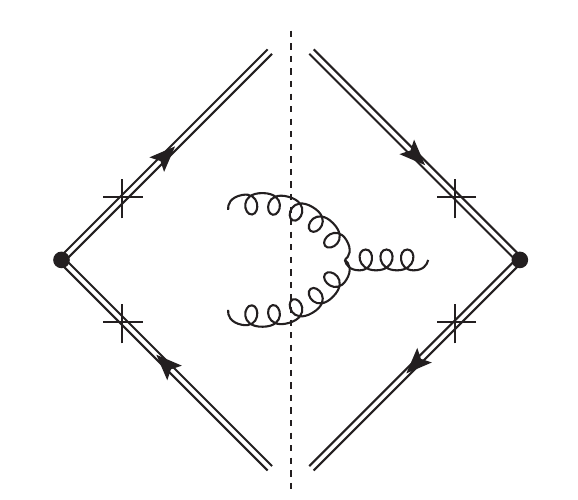}\\
(a)	&	(b)	&	(c)\\
\includegraphics[width=4cm]{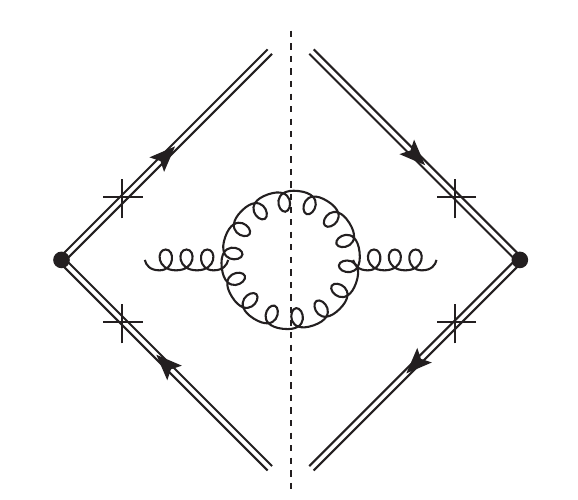}&
\includegraphics[width=4cm]{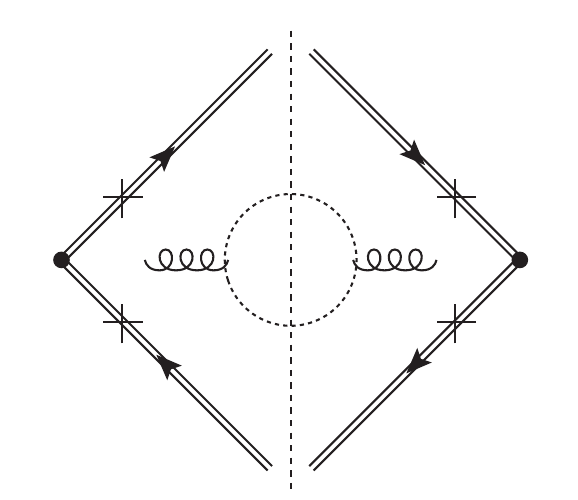}&
\includegraphics[width=4cm]{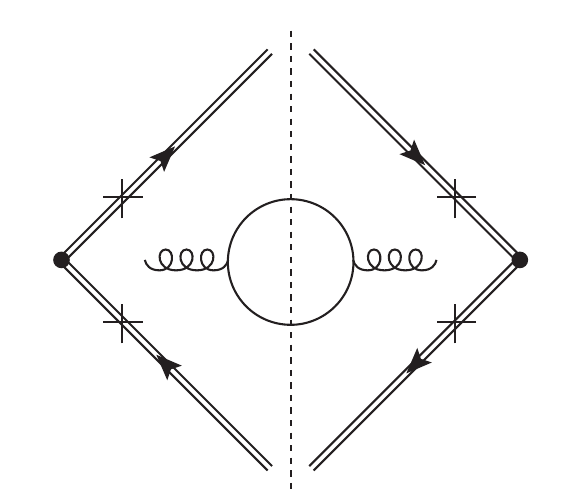}\\
(d)	&	(e)	&	(f)
\end{tabular}
\caption{All soft double real emission (real-real) diagrams contributing at NNLO. The endpoints of the gluon lines can connect to any of the crossed points on their side of the cut. The diagrams are divided into five groups: three double gluon groups  $\mathcal{I}$ = (a), $\mathcal{T}$ = (b)+(c) and $\mathcal{G}$ = (d); one with ghosts $\mathcal{H}$ = (e); and one with $n_f$ massless fermions $\mathcal{Q}$ = (f). Their color structure can be read off \Tab{color} and their (unintegrated) amplitudes are given in \mycite{Hornig:2011iu}.}
\label{fig:rr}
\end{center}
\end{figure}

The two-loop diagrams for the TMD soft function are actually the same as for the dijet ($e^+e^-$ hemisphere) soft functions in \mycites{Kelley:2011ng,Monni:2011gb,Hornig:2011iu}.
Only the measurement (delta) function differs.
In \mycite{Hornig:2011iu} the explicit expressions for the unintegrated amplitudes of the double real emission diagrams shown in \fig{rr} are given in Feynman gauge.
We can readily use these amplitudes for our calculation. We just have to implement our measurement and rapidity regulator according to \subsecs{ptReg}{regulator} and finally integrate as described below. 
Similar to \mycite{Hornig:2011iu}, we divide the double real emission diagrams into five groups as displayed in \fig{rr}. 
With the right choice of integration variables, all the real-real diagrams can be expressed through the same family of two-loop integrals. They can therefore be calculated with the same method, which we now describe in detail.

The two integrations are over the $d$-dimensional real particle momenta $k_1$ and $k_2$. 
With our regulator and measurement, it turns out to be convenient to use the integration variables $\ell=k_1$ and $k=k_1+k_2$. 
The phase space integration then takes the generic form
\begin{align}\label{eq:RRintegral}
\int\!\frac{\df^d\ell}{(2\pi)^d}\int\!\frac{\df^dk}{(2\pi)^d} \, \abs{k^--k^+}^{-\eta}
\,\delta^+(\ell^2)\,\delta^+\left((k-\ell)^2\right)\,\delta^{(2)}(\mathbf{p}_{\perp}-\mathbf{k}_\perp)
\,\mathcal{A}(\ell,k)\,,
\end{align}
where the two $\delta^+(k_i^2) \equiv \theta(k_i^0) \delta(k_i^2)$ originate from cutting the propagators of the two real particles. 
The expressions for the unintegrated amplitudes $\mathcal{A}_i$ for $i=\{\mathcal{I}, \mathcal{T}, \mathcal{G}, \mathcal{H}, \mathcal{Q}\}$ can directly be taken from appendix B of \mycite{Hornig:2011iu}.

Considering the $k$ integral first, 
we find a set of convenient integration variables with the $d-2$ dimensional $\mathbf{k}_\perp$ and
\begin{align}
y&=\frac{\mathbf{k}_\perp^2}{k^-k^+}\geq 0\,,\\
v&=k^--k^+.
\end{align}
A major advantage of this variable choice is that only the regulator factor $\abs{v}^{-\eta}$ depends on $v$.
Hence the integration over $k$'s light-cone components reduces to
\begin{align}\label{eq:lcintegration}
  \int \!\!\df k^0\int \!\!\df k^3 \, \abs{k^--k^+}^{-\eta}&=\int_0^1 \df y\int_{-\infty}^\infty \df v \abs{v}^{-\eta}\frac{\mathbf{k}_\perp^2}{y^{3/2}\sqrt{yv^2+4\mathbf{k}_\perp^2}}\nn\\
  &=2^{-\eta}\,
  \pi^{-1/2}\Gamma\Big(\frac{1}{2}-\frac{\eta}{2}\Big) \, \Gamma\Big(\frac{\eta}{2}\Big) \,
  (\mathbf{k}_\perp^2)^{1-\frac{\eta}{2}}\!\int_0^1 \!\df y\, y^{-2+\frac{\eta}{2}}
  \,.
\end{align}

To perform the $\ell$ integral in \eq{RRintegral}, we boost to the rest frame of $k$, where the involved vectors can be parametrized as
\begin{align}
k^\mu&=|\mathbf{k}_\perp|\sqrt{\frac{1-y}{y}}(1,\boldsymbol{0})\,,\\
\bar{n}&=\frac{k^-}{|\mathbf{k}_\perp|}\sqrt{\frac{y}{1-y}}(1,0,0,1)\,,\\
n&=\frac{|\mathbf{k}_\perp|}{k^-}\frac{1}{\sqrt{y(1-y)}}(1,0,-2\sqrt{y(1-y)},2y-1)\,,\\
\ell&=\frac{|\mathbf{k}_\perp|}{2}\sqrt{\frac{1-y}{y}}(1,\ldots,\sin\theta_1\cos\theta_2,\cos\theta_1)\,,
\end{align}
where the dots in $\ell$ denote the remaining $d-3$ components of the unit vector in spherical coordinates.
Now we can express the amplitude in terms of $|\mathbf{k}_\perp|$, $k^-$ and $y$, as well as the angles $\theta_i$.
All angle dependent scalar products that can appear in the amplitudes are then
\begin{align}
  	\bar{n}\cdot\ell&=k^-\frac{1-\cos\theta_1}{2}\equiv k^-D_1 \,, \\
  	n\cdot\ell&= \frac{\mathbf{k}_\perp^2}{k^-y}\frac{1}{2}\left(1+2\sqrt{y(1-y)}\sin\theta_1\cos\theta_2+(2y-1)\cos\theta_1\right)\equiv	\frac{\mathbf{k}_\perp^2}{k^-y}D_2 \,,	\\
  	\bar{n}\cdot(k-\ell)&=k^-(1-D_1)\equiv k^-D_3 \,,	\\
  	n\cdot(k-\ell)&=\frac{\mathbf{k}_\perp^2}{k^-y}(1-D_2)\equiv\frac{\mathbf{k}_\perp^2}{k^-y}D_4 \,.
\end{align}
Note that none of these factors depends on the other $d-3$ spherical coordinates.
Hence, the corresponding angular integral can be performed trivially.
It is also convenient to define
\begin{align}
  D_5 &=y \,,\\
  D_6 &=1-y \,.
\end{align}

Implementing the above parameterizations in \eq{RRintegral}, the $\mathbf{k}_\perp$ dependence is power-like and we can easily perform its integral:
\begin{align}
  \int \df^{d-2}\mathbf{k}_\perp(\mathbf{k}_\perp^2)^{-1-\epsilon-\frac{\eta}{2}}\delta^{(2)}(\mathbf{p}_{\perp}-\mathbf{k}_\perp)=\pi^{-\e}\frac{\Gamma\left(1+2\e+\frac{\eta}{2}\right)}{\Gamma\left(1+\e+\frac{\eta}{2}\right)} (\pt^2)^{-1-2\epsilon-\frac{\eta}{2}}\,.
\end{align}
The $\ell^0$ and $\abs{\boldsymbol{\ell}}$ integrals can be done with the delta functions in \eq{RRintegral} and we finally arrive at integrals of the form 
\begin{align}\label{eq:IRRdef}
  I^{RR}(a_1,a_2,a_3,a_4,a_5,a_6) &\equiv\int_0^1 \! \df y \, \frac{1}{2\pi}\frac{\Gamma^2(1-\epsilon)}{\Gamma(1-2\epsilon)}\int_0^\pi \! \df\theta_1\sin^{1-2\epsilon}\theta_1\int_0^\pi \! \df\theta_2 \sin^{-2\epsilon}\!\theta_2 \nn\\
  & \times D_5^{\epsilon+\tfrac{\eta}{2}} \, D_6^{-\epsilon} \, \prod_{i=1}^6D_i^{-a_i}\,.
\end{align}
This integral family is studied in \mycites{Gehrmann:2014yya,Thomasthesis} and we can use their results for the relevant integrals in our case:
\begin{align}
I^{RR}(a_1,0,a_3,0,a_5,a_6)&=\frac{\Gamma(1-a_5+\e+\tfrac{\eta}{2})\Gamma(1-a_6-\e)}{\Gamma(2-a_5-a_6+\tfrac{\eta}{2})}\frac{\Gamma(1-\epsilon-a_1)\Gamma(1-\epsilon-a_3)}{\Gamma(2-2\epsilon-a_1-a_3)}\,,\label{eq:Isol1}\\
I^{RR}(0,a_2,a_3,0,a_5,a_6)&=\frac{\Gamma(1-\epsilon-a_2)\Gamma(1-\epsilon-a_3)}{\Gamma(2-a_2-a_3-2\epsilon)}\frac{\Gamma(1-a_6-\e)\Gamma(2-a_2-a_3-a_5+\tfrac{\eta}{2})}{\Gamma(3-a_2-a_3-a_5-a_6-\e+\tfrac{\eta}{2})}\nn\\
&\times{_3F_2}\Bigg(\!\!\!\!
\begin{array}{c} 1-a_6-\e,1-\epsilon-a_2,1-\epsilon-a_3\\ 3-a_2-a_3-a_5-a_6-\epsilon+\tfrac{\eta}{2},1-\epsilon
\end{array}
\!\!;1\Bigg),
\label{eq:Isol3}\\
I^{RR}(a_1,a_2,0,0,a_5,a_6)&=\frac{\Gamma(1-a_5+\e+\tfrac{\eta}{2})\Gamma(1-a_6-\e)\Gamma(1-\epsilon-a_1)\Gamma(1-\epsilon-a_2)}{\Gamma(2-a_1-a_2-a_6-\e)\Gamma(2-a_1-a_2-2\epsilon)}\nn\\
&\times{_3F_2}\Bigg(\!\!\!\!
\begin{array}{c} 
a_1,a_2,1-a_5+\e+\tfrac{\eta}{2}\\ 2-a_5-a_6+\tfrac{\eta}{2},1-\e
\end{array}
\!\!;1\Bigg).
\label{eq:iRR1}
\end{align}
Note that \eq{iRR1} is not valid for $1\leq a_1+a_2\leq 2$.\footnote{The $_3F_2$ hypergeometric function arises from an integral over a $_2F_1$. For the values $1\leq a_1+a_2\leq 2$ one integrates over a singularity and the $_3F_2$ is therefore not well defined. Using an Euler transformation on the $_2F_1$, the singular part can be extracted to arrive at a well defined integral representation of a $_3F_2$.} 
In this case we must instead use
\begin{align}
\label{eq:iRR2}
I^{RR}(a_1,a_2,0,0,a_5,a_6)&=\frac{\Gamma(1-a_5+\e+\tfrac{\eta}{2})\Gamma(2-a_1-a_2-a_6-2\e)}{\Gamma(3-a_1-a_2-a_5-a_6-\epsilon+\tfrac{\eta}{2})} \\
&\hspace{-3cm}\times\frac{\Gamma(1-\epsilon-a_1)\Gamma(1-\epsilon-a_2)}{\Gamma(2-a_1-a_2-2\epsilon)} \;
{_3F_2}\Bigg(\!\!\!\!
\begin{array}{c} 
1-a_5+\e+\tfrac{\eta}{2},1-\epsilon-a_1,1-\epsilon-a_2\\ 3-a_1-a_2-a_5-a_6+\tfrac{\eta}{2}-\epsilon,1-\epsilon
\end{array}
\!\!;1\Bigg).\nn
\end{align}

Now, matching the amplitudes onto the $I^{RR}$ functions in \eqss{Isol1}{iRR2} gives the final results for the double real diagram groups defined in \fig{rr}:
\begin{align}
\mathcal{I}_{C_FC_A}&=-C_FC_AK_{\mathrm{RR}}\left[2I^{RR}(1,1,0,0,0,0)+I^{RR}(0,1,1,0,0,0)\right],\label{eq:rrICF}\\
\mathcal{T}&=C_FC_AK_{\mathrm{RR}}\left[2I^{RR}(-1,0,1,0,0,1)+I^{RR}(0,1,1,0,0,1)\right],\\
\mathcal{G}+\mathcal{H}&= C_FC_AK_{\mathrm{RR}} 
\left[
(1-\epsilon) \left\{I^{RR}(-2,0,0,0,0,2)-2I^{RR}(-1,-1,0,0,0,2)\right.\right.\nn\\
&\left.\left.+I^{RR}(0,-2,0,0,0,2) \right\}
-2I^{RR}(0,0,0,0,0,1)
\right],\\
\mathcal{Q}&=2C_FT_Fn_fK_{\mathrm{RR}}\left[-I^{RR}(-2,0,0,0,0,2)+2I^{RR}(-1,-1,0,0,0,2)\right.\nn\\
&\left.-I^{RR}(0,-2,0,0,0,2)+I^{RR}(0,0,0,0,0,1)\right]\,.\label{eq:rrQ}
\end{align}
The common prefactor is
\begin{align}
K_{\mathrm{RR}}=4a_s^2\frac{e^{2\e\gamma_E}\mu^{4\epsilon}\nu^\eta w^2}{\pi^{3/2}}\frac{\Gamma\left(\frac{1}{2}-\frac{\eta}{2}\right)\Gamma\left(\frac{\eta}{2}\right)\Gamma\left(1+2\e+\frac{\eta}{2}\right)}{2^\eta\Gamma(1-\epsilon)\Gamma\left(1+\e+\frac{\eta}{2}\right)}\, (\pt^2)^{-1-2\epsilon-\frac{\eta}{2}}
\,.
\end{align}
Of the eight distinct $I^{RR}$ functions in \eqss{rrICF}{rrQ}, five reduce to combinations of ordinary gamma functions and are thus straightforwardly expanded in $\eta$ and $\epsilon$.
The remaining three $I^{RR}$ functions contain nontrivial hypergeometric functions which are expanded using the {\tt HypExp} package~\cite{Huber:2005yg,Huber:2007dx}. We list the corresponding results in \app{expansion}.
\subsubsection{Non-Abelian exponentiation}
\label{subsec:exponentiation}

In the calculation of the soft function we have so far neglected any $C_F^2$ pieces in the diagrams. 
Instead of computing it explicitly, we can obtain the $C_F^2$ part from the non-Abelian exponentiation theorem \cite{Frenkel:1984pz,Gatheral:1983cz}. 
According to this theorem, the soft function takes the form of an exponential of webs in position space. 
We therefore write the bare soft function as
\begin{align}\label{eq:Sexp}
  S_\bare^q(\bb)&=\frac{1}{(2\pi)^2}\exp\bigg[\sum_{n=1}^\infty {s^q}^{(n)}(\bb)\, a_s^n\bigg]\nn\\
&=\frac{1}{(2\pi)^2}\bigg[1+{s^q}^{(1)}(\bb)\,a_s+\Big(\frac{1}{2} \big[{s^q}^{(1)}(\bb)\big]^2+{s^q}^{(2)}(\bb)\Big)\,a_s^2+\mathcal{O}(a_s^3)\bigg],
\end{align}
where the factor of $1/(2\pi)^2$ is a normalization factor. 
We can identify the $\mathcal{O}(a_s)$ term in the exponent with the Fourier transform of the NLO bare TMD soft function computed in \eq{nlobare}. 

The non-Abelian exponentiation theorem states that ${s^q}^{(1)}(\mathbf{b})$ is purely $C_F$, while ${s^q}^{(2)}(\mathbf{b})$ only contains the color factors $C_FC_A$ and $C_F T_F n_f$. 
Hence, the total $C_F^2$ part of the soft function is equal to the $[{s^q}^{(1)}(\mathbf{b})]^2$ term in \eq{Sexp}. 
This allows us to obtain the $C_F^2$ piece by simply squaring the NLO result. 
Fourier transformation to momentum space turns the square into a convolution, and the full $C_F^2$ part of the soft function is
\begin{align}\label{eq:S_cf2}
  {S^q_\bare}^\two(\pt) \Big|_{C_F^2} = \mathcal{F}\left[\frac{1}{(2\pi)^2}\frac{1}{2}{s^q}^{(1)}(\bb)^2\right]=\frac{(2\pi)^2}{2}
  {S^q_\bare}^\one(\pt)\otimes_\perp {S^q_\bare}^\one(\pt)\,. 
\end{align}	

It is straightforward to see that the total $C_F^2$ part comes exclusively from the $\mathcal{I}$ diagrams in \fig{rr}. 
With a rapidity regulator for each real gluon momentum, the $C_F^2$ piece of $\mathcal{I}$ takes the form
\begin{align}\label{eq:ICF2}
	\mathcal{I}_{C_F^2}(\pt)&=8\times(4\pi)^2C_F^2\alpha_s^2w^4\nu^{2\eta}\mu^{4\e}\left(\frac{e^{\gamma_E}}{4\pi}\right)^{2\e}\nn\\
	&\times\int\frac{\df^dk_1}{(2\pi)^d}\int\frac{\df^dk_2}{(2\pi)^d}\,|k_1^--k_1^+|^{-\eta}\,|k_2^--k_2^+|^{-\eta}
	\,\frac{(2\pi)\,\delta^+(k_1^2)\,(2\pi)\,\delta^+(k_2^2)}{k_1^+k_1^-k_2^+k_2^-}\nn\\
	&\times\delta^{(2)}({\mathbf{k}_1}_\perp+{\mathbf{k}_2}_\perp-\pt)\,.
\end{align}
Simply performing the convolution integral in \eq{S_cf2} we find
\begin{align}
 \mathcal{I}_{C_F^2}(\pt) = {S^q_\bare}^\two(\pt) \Big|_{C_F^2} \,.
\end{align}
Consequently the (non-web) $C_F^2$ piece of the real-virtual diagrams (b) and (c) in \fig{rv} must be zero.
In fact, one can see this already at the integrand level. In the sum of the diagrams the virtual part factorizes due to the eikonal identity and gives a vanishing scaleless integral as an overall factor. 
With real gluon momentum $\ell$ and virtual gluon momentum $k$, the two-loop integral explicitly evaluates to
\begin{align}\label{eq:RCF2}
	\mathcal{R}_{C_F^2}(\pt)&\propto C_F^2\int\frac{\df^d\ell}{(2\pi)^d}\,|\ell^--\ell^+|^{-\eta}\frac{(2\pi)\,\delta^+(\ell^2)\,\delta^{(2)}(\pt-\boldsymbol{\ell}_\perp)}{\ell^+\ell^-}\nn\\
	&\times\int\frac{\df^dk}{(2\pi)^d}\,|k^--k^+|^{-\eta}\frac{1}{k^+k^-k^2}=0\,.
\end{align} 
The calculations in \eqss{ICF2}{RCF2} explicitly show that the $\eta$-regulator preserves non-Abelian exponentiation at two loops. We stress that the separate regulation of each web (in this case gluon) momentum is crucial for this.
Likewise the non-Abelian exponentiation holds for the RRG renormalized TMD soft function.

\subsection{Results in momentum space}
\label{subsec:results}

The final result for the bare NNLO soft function in the fundamental representation is
\begin{align}\label{eq:Sbare}
S_\bare^q
&= \delta^\two(\pt) + S^q_{\bare,\rm 1loop}
+\mathcal{I}_{C_F^2}
+\mathcal{R}_{C_FC_A}+\mathcal{I}_{C_FC_A}+\mathcal{T}+\mathcal{G}+\mathcal{H}+\mathcal{Q}+\ordo{a_s^3}.
\end{align}
All pieces have been calculated in the previous sections. 
The bare one-loop result can be found in \eq{nlobare} and in \mycite{Chiu:2012ir}.
The $C_FC_A$ part of the NNLO real-virtual diagrams, $\mathcal{R}_{C_FC_A}$, is written in \eq{rCFCAbare}.
As shown in \subsec{exponentiation}, we can use the non-Abelian exponentiation theorem to express the $C_F^2$ part ($=\mathcal{I}_{C_F^2}$) in terms of the NLO expression in \eq{nlobare}.
The double real terms, $\mathcal{I}_{C_FC_A},\mathcal{T},\mathcal{G},\mathcal{H}$ and $\mathcal{Q}$, are given in \eqss{rrICF}{rrQ}.

For the expansion of the bare soft function in $\epsilon$ and $\eta$ we make use of \eq{plusidentity}. The expanded momentum space results will therefore involve the plus distributions $\plus{n}(\pt,\mu)$ defined in \eqs{plusDef1}{plusDef2}.
We give tables for Fourier transformations and the relevant convolutions among the plus distributions in \app{plusDistributions}.
In the following we will usually suppress the arguments of the $\plus{n}$.

After expanding, we renormalize the soft functions according to \eq{Sbare2} and find the NLO counterterm and soft function coefficients in \eq{S_coef}:
\begin{align}\label{eq:ZNLO}
{Z_S^i}^{(1)}(\pt,\mu,\nu)&=4\pi^2\Gamma_0^iw^2\left[\frac{1}{\eta}\left(4\plus{0}-\frac{2}{\epsilon}\plus{-1}+\ordo{\e}\right)+\left(\frac{1}{\epsilon^2}+\frac{2}{\epsilon}\ln\frac{\mu}{\nu}\right)\plus{-1}\right]\,\\
{S^i}^{(1)}(\pt,\mu,\nu)&=S_1^i\plus{-1}+2\Gamma_0^i\left(\plus{1}-2\plus{0}\ln\frac{\mu}{\nu}\right)+\mathcal{O}\left(\eta\right)+\mathcal{O}\left(\epsilon\,\eta^0\right)\,.
\label{eq:SrNLO}
\end{align}
They agree with the ones of \mycite{Chiu:2012ir} (for $i=g$).%
\footnote{Up to misprints in eq.~(5.62) and eq.~(5.63) in the arXiv and journal version of their article, respectively.}
Similarly, we obtain the NNLO coefficients
\begin{align}
	{Z_S^i}^{(2)}(\pt,\mu,\nu)&=\frac{\pi^2w^4(\Gamma_0^i)^2}{\eta^2}\left[\Big(\frac{4\pi^2}{3}+\frac{8}{\e^2}\Big)\plus{-1}-\frac{32}{\e}\plus{0}-64\plus{1}+\ordo{\e}\right]\nn\\
	&\hspace{-4 ex}+\frac{\pi^2w^2}{\eta}\Bigg\{4{\gamma_\nu^i}_1\plus{-1}+\beta_0\Gamma_0^i\Big(16\plus{1}+\frac{4}{\e^2}\plus{-1}\Big)+\Gamma_1^i\Big(16\plus{0}-\frac{4}{\e}\plus{-1}\Big)\nn\\
	&\hspace{-4 ex}+w^2(\Gamma_0^i)^2\Bigg[\left(\frac{8\zeta(3)}{3}-\frac{8}{\e^3}-\frac{2\pi^2}{3\e}-\Big(\frac{4\pi^2}{3}+\frac{16}{\e^2}\Big)\Log\right)\plus{-1}+8\plus{2}\nn\\
	&\hspace{-4 ex}+\Big(\frac{16}{\e}+32\Log\Big)\plus{1}+\Big(\frac{4\pi^2}{3}+\frac{16}{\e^2}+\frac{32}{\e}\Log\Big)\plus{0}\Bigg]+\ordo{\e}\Bigg\}\nn\\
	&\hspace{-4 ex}+\pi^2w^2\plus{-1}\Bigg[\Gamma_1^i\Big(\frac{1}{\e^2}+\frac{4}{\e}\Log\Big)+w^2(\Gamma_0^i)^2\Big(\frac{2}{\e^4}+\frac{8}{\e^3}\Log+\frac{8}{\e^2}\Logs\Big)\nn\\
	&\hspace{-4 ex}+\frac{{\gamma_S^i}_1}{\e}-\beta_0\Gamma_0^i\Big(\frac{3}{\e^3}+\frac{4}{\e^2}\Log\Big)\Bigg]
	\,,\!
\label{eq:Z2}\\[2 ex]
\label{eq:S2}
{S^i}^{(2)}(\pt,\mu,\nu)&=-(\Gamma_0^i)^2\plus{3}+\left(2\beta_0\Gamma_0^i+6(\Gamma_0^i)^2\Log\right)\plus{2}\nn\\
&+\left[2\Gamma_1^i-8(\Gamma_0^i)^2\Logs+\Gamma_0^i\left(2S_1^i-4\beta_0\Log\right)\right]\plus{1}\nn\\
&+\left[-{\gamma_\nu^i}_1-{\gamma_S^i}_1-2\beta_0S_1^i-4\left(S_1^i\Gamma_0^i+\Gamma_1^i\right)\Log+4(\Gamma_0^i)^2\zeta(3)\right]\plus{0}\nn\\
&+\left[S_2^i+\frac{4}{3}\beta_0\Gamma_0^i\zeta(3)+\left(4(\Gamma_0^i)^2\zeta(3)-{\gamma_\nu^i}_1\right)\Log\right]\plus{-1}\nn\\
&+\ordo{\eta}+\ordo{\e\,\eta^0}\,.
\end{align}
Here we have generalized our results to an arbitrary color representation according to Casimir scaling and parametrized them in terms of the anomalous dimension and integration constants introduced in \subsec{logStructure} for conciseness.
The explicit values for the constants ${\gamma_S^i}_n$, ${\gamma_\nu^i}_n$ and $S_n^i$ extracted from the above results are given below in \eqss{gammaconst1}{Sconst2}.

In ${S^i}$ we have set $w=1$. In the renormalization factor ${Z_S^i}$ we must however keep the $\eta$-dependent $w$
as it is crucial for the derivation of the RRGE, see \eq{wBeta}.
With \eqs{gamma_mu_def}{gamma_nu_def} we thus find for the anomalous dimensions
\begin{align}
\label{eq:gamma_S_pt_0}
{\gamma_{S,\mu}^i}^{\!\!\!\!\!(0)}(\mu,\nu) &= 4\Gamma_0^i\ln\frac{\mu}{\nu}\,,\\
{\gamma_{S,\mu}^i}^{\!\!\!\!\!(1)}(\mu,\nu) &= 4\Gamma_1^i\ln\frac{\mu}{\nu} + {\gamma_S^i}_1\,,\label{eq:gamma_S_pt_1}\\
{\gamma_\nu^i}^{(0)}(\pt,\mu)&=4 (2\pi)^2\plus{0} \Gamma_0^i \,,\label{eq:gNuPt1}\\
{\gamma_\nu^i}^{(1)}(\pt,\mu)&=4 (2\pi)^2\left(\plus{0} \Gamma_1^i +\plus{1}\Gamma_0^i\beta_0 \right) + (2\pi)^2\plus{-1} {\gamma_\nu^i}_1
\,.
\label{eq:gNuPt2}
\end{align}
Note that the subleading terms in $\e$ and $\eta$ in \eqs{ZNLO}{SrNLO} are crucial for the derivation of \eqs{Z2}{S2}.
The subleading $\e$ and $\eta$ terms in \eqs{Z2}{S2} will in turn be relevant at higher orders.
Such terms are of course also present in the position space results and the results for the beam functions given below. For the sake of brevity we will suppress them in the following.

We note, that our momentum space results depend on the renormalization scales $\mu$ and $\nu$ only through $\plus{n}$, $\ln \frac{\mu}{\nu}$ and $a_s$.
Correspondingly, the position space results depend on $\mu$ and $\nu$ only through $L_b$, $\ln \frac{\mu}{\nu}$ and $a_s$. The scale dependence is dictated by the RGEs and RRGEs. In \subsec{logStructure}, we have made this explicit in terms of the recursion relations \eqs{nurecurrence}{Srecurrence}. They are in perfect agreement with our results for the soft function and its $\nu$ anomalous dimension up to two loops.

\subsection{Results in position space}
\label{subsec:bspace}

Fourier transforming the $\pt$-dependent soft function in the previous section to impact parameter space or using the recurrence relation in \eq{Srecurrence} gives
\begin{align}\label{eq:Sbspace0}
{S^i}^{(0)}(\bb,\mu,\nu)&=\frac{1}{(2\pi)^2}S^i_0,\\
{S^i}^{(1)}(\bb,\mu,\nu)&=\frac{1}{(2\pi)^2}\left[S^i_1+2\Gamma_0^i \ln\frac{\mu}{\nu}L_b-\frac{\Gamma_0^i}{2}L_b^2\right],
\label{eq:Sbspace1}\\
{S^i}^{(2)}(\bb,\mu,\nu)&=\frac{1}{(2\pi)^2}\Bigg[S^i_2-{\gamma_\nu^i}_1\ln\frac{\mu}{\nu} \nn
\label{eq:Sbspace2}\\
&+\left(\frac{{\gamma_\nu^i}_1}{2}+\frac{{\gamma_S^i}_1}{2}+\beta_0 S^i_1+2(S^i_1\Gamma_0^i+\Gamma_1^i)\ln\frac{\mu}{\nu}\right)L_b\nn\\
&+\left(-\frac{S^i_1 \Gamma_0^i}{2}-\frac{\Gamma_1^i}{2}+\beta_0\Gamma_0^i\ln\frac{\mu}{\nu}+2(\Gamma_0^i)^2\ln^2\frac{\mu}{\nu}\right)L_b^2\nn\\
&+\left(-\frac{\beta_0\Gamma_0^i}{3}-{\Gamma_0^i}^2\ln\frac{\mu}{\nu}\right)L_b^3+\frac{{\Gamma_0^i}^2}{8}L_b^4\Bigg].
\end{align}
Similarly, one can Fourier transform the $\nu$ anomalous dimension coefficients in \eqs{gNuPt1}{gNuPt2} or use the recurrence relation in \eq{nurecurrence} to arrive at
\begin{align}\label{eq:gNuBspace}
{\gamma_\nu^i}^{(0)}(\bb,\mu)&=-2\Gamma_0^iL_b+{\gamma_\nu^i}_0 \,,\\
{\gamma_\nu^i}^{(1)}(\bb,\mu)&=-\beta_0\Gamma_0^i L_b^2-2\Gamma_1^iL_b+{\gamma_\nu^i}_1 \,.
\end{align}
The $\mu$ anomalous dimension coefficients according to \eqs{pTgamma_Smu}{gamma_S_i} have already been written in eqs.~\eqref{eq:gamma_S_pt_0},\eqref{eq:gamma_S_pt_1}.
The constants that are not predicted by the RGE and RRGE are fixed by our explicit calculation of the soft function to
\begin{align}
\label{eq:gammaconst1}
{\gamma_S^i}_0&={\gamma_\nu^i}_0=0\,,\\
\label{eq:gammaconst2}
{\gamma_S^i}_1&=\mathcal{C}^i\bigg[C_A\bigg(\frac{128}{9}-56\zeta(3)\bigg)+\beta_0\bigg(\frac{112}{9}-\frac{2\pi^2}{3}\bigg)\bigg]\,,\\
{\gamma_\nu^i}_1&=\mathcal{C}^i\bigg[-C_A\bigg(\frac{128}{9}-56\zeta(3)\bigg)-\beta_0\frac{112}{9} \bigg]\,,
\label{eq:gammaconst3}\\
S^q_0&=S^g_0=1\,,\\
S^i_1&=-\frac{\mathcal{C}^i\pi^2}{3}\,,\\
\label{eq:Sconst2}
S^i_2&=\frac{{\mathcal{C}^i}^2\pi^4}{18}+\mathcal{C}^i\bigg[C_A\bigg(\frac{208}{27}-\frac{2\pi^2}{3}+\frac{\pi^4}{9}\bigg)+\beta_0\bigg(\frac{164}{27}-\frac{5\pi^2}{6}-\frac{14\zeta(3)}{3}\bigg)\bigg]
\,,
\end{align}
where we have introduced the color factor $\mathcal{C}^i$, which equals $C_F$ ($C_A$) for $i=q$ ($i=g$).

We note the striking similarity of the (non-cusp) constants ${\gamma_S^i}_1$ and ${\gamma_\nu^i}_1$.
Given the scheme dependence of ${\gamma_S^i}_1$, we refrain however from investigating this further at the time.

\section{Comparison to the literature}
\label{sec:comparison}

Several different formulations of the SCET TMD factorization theorem can be found in the literature.
They differ by the treatment of the rapidity divergences, i.e.\ the way they are regulated,
and how the individual soft and collinear functions are defined.
Of course formulations in momentum as well as in position space are possible and directly related by Fourier transformation.
While the combination of the TMD soft and beam functions is unambiguous, the three parts individually are scheme-dependent.
In our case, this is reflected by the dependence on the unphysical scale $\nu$.
Varying $\nu$ allows a systematic quantification of effects from higher order terms related to the rapidity logarithms and therefore contributes to a reliable estimate of the overall perturbative uncertainty of the cross section.

In \subsec{rel_BN} we study the relation of our scheme to the one of \mycites{Becher:2010tm,Becher:2012yn}.
The NNLO factorization ingredients for the latter have been calculated in \mycites{Gehrmann:2012ze,Gehrmann:2014yya}.
Recently, the authors of \mycite{Echevarria:2015byo} have computed the soft function in the scheme of \mycite{GarciaEchevarria:2011rb} to NNLO. The result still contains explicit poles in $\eps$ and their rapidity regulator. 
The corresponding beam functions are only known to NLO so far.%
\footnote{Note that the NNLO calculations in \mycites{Gehrmann:2012ze,Gehrmann:2014yya,Echevarria:2015byo} all have been performed with slightly modified regulators, which are better suited for higher loop calculations, as compared to the initial papers.}

In \mycites{Becher:2010tm,Becher:2012yn} the bare ($\bb$ space) soft function is, by construction and to all orders in perturbation theory, identical to one. 
The analog to the renormalized/resummed soft function is absorbed into the 
collinear functions (TMDPDFs) and the `collinear anomaly' exponential as we 
will see below.
In \mycites{Gehrmann:2012ze,Gehrmann:2014yya} the corresponding TMDPDFs and the collinear anomaly exponent are determined to NNLO.
Identifying the exact relation to our scheme 
together with our NNLO results for the soft function
allows us to extract the NNLO expressions for our beam functions.
We will work this out in detail below.
The results for the rapidity renormalized NNLO beam functions we present in section \ref{sec:beam}.

TMD distributions have actually first been considered in traditional QCD. Again, the final result for the differential cross section should agree with the SCET predictions (up to higher order terms). 
The relation between the SCET scheme we are comparing to in \subsec{rel_BN} and
the direct QCD approach of 
\mycites{Collins:1984kg,deFlorian:2000pr,deFlorian:2001zd,Bozzi:2005wk, 
Catani:2010pd,deFlorian:2011xf,Catani:2013tia, Catani:2011kr, Catani:2012qa} 
has been discussed in \mycites{Becher:2010tm,Gehrmann:2014yya}.

\subsection{Extraction of the TMD beam functions in the RRG scheme}
\label{subsec:rel_BN}

In \subsec{logStructure}, we have obtained recurrence relations that can be used to derive the logarithmic structure of the soft and beam functions to all orders. We have also computed the complete NNLO soft function and thus know by virtue of \eq{murelation} all anomalous dimensions at this order.
Hence, the pieces we are missing are the boundary terms $I_{ij}^{(2)}(z)$ and $J_{gj}^{(2)}(z)$ in the beam function matching coefficients, see \eq{Irecurrence}.

We now show how to determine them from the results for the NNLO TMD beam functions in \mycites{Gehrmann:2012ze,Gehrmann:2014yya}, which are based on the framework of \mycites{Becher:2010tm,Becher:2012yn}.
The comparison of the two formalisms is most conveniently performed in position space, where the TMD convolutions turn into ordinary products and the $\plus{n}$ distributions into logarithms $L_b$.
The translation between position and momentum space can easily be carried out with the help of tables~\ref{tab:fourier} and \ref{tab:fourier2}.

To perform the comparison on the level of perturbative functions,
we deconvolve the universal PDFs $f_i$
from the collinear functions and divide out the common hard function.
The equality of the (perturbative) physical cross section then implies
\begin{align}
(2\pi)^6 \,
S^k(\bb,\mu,\nu)\; \mathcal{I}_{ki}(z_1,\bb,\mu,\omega_+/\nu)\; \mathcal{I}_{\bar{k}j}(z_2,\bb,\mu,\omega_-/\nu) 
\nn\\
=\left(\frac{\mathbf{b}^2Q^2}{4e^{-2\gamma_E}}\right)^{-F_{k}(\bb,\mu)} \, I_{k/i}(z_1,\bb,\mu) \, I_{\bar{k}/j}(z_2,\bb,\mu)\,,
\label{eq:IIS}
\end{align}
where $k,i,j \in \{g,q,\bar{q},q'\}$, $\bar{g}=g$.%
\footnote{The names of the variables in \mycites{Gehrmann:2012ze,Gehrmann:2014yya} have been translated to our notation as $x_T^2=\bb^2$ and $L_\perp = L_b$. They also use a different definition of $P^{(n)}_{ij}(z)$ which is related to ours by adding a factor $2^{-n}$.}
There are additional equations for $k=g$,
where one or both of the pairs $(\mathcal{I}_{gl},I_{g/l})$ are exchanged for $(\mathcal{J}_{gl},I'_{g/l})$ associated with the other tensor structure in \eq{Bg}.
These cases are understood in the discussion below.

The expression in the first line of \eq{IIS} are the combined soft and collinear parts from the renormalized version of our factorization formulas in \eqs{factq}{factg}. The second line corresponds to the equivalent expressions in the scheme of \mycites{Becher:2010tm,Becher:2012yn}, where
the first term is a resummed expression generated by the 'collinear anomaly' with the respective coefficient $F_k$ and the $I_{i/j}$ are the TMDPDFs.
The details can be found in section 2.2 of \mycite{Gehrmann:2014yya}. Note especially eq.~(2.1) therein.
The factor $(2\pi)^6$ arises because the normalizations of the three functions on both sides of \eq{IIS} differ by a factor of $(2\pi)^2$ each.

Strictly speaking \eq{IIS} only holds once both sides are expanded to a fixed order in $\alpha_s$ (or for the hypothetical exact all-order solutions of beam and soft functions).
If we include the resummation factors, according to \eq{resum_formal_sol}, and the analog for the $I_{i/j}$ at a finite logarithmic order, in general the two sides of \eq{IIS} will disagree by terms beyond that order.
The reason is related to the resummation of the rapidity logarithms.
The lack of the $\nu$ argument on the RHS of \eq{IIS} indicates that for resummed beam and soft functions exact agreement of both sides is only achieved for a specific choice of the renormalization scales $\nu_B^\pm$ and $\nu_S$, which turn out to be the canonical scales in \eq{defaultscales}.
Note that the $\nu$ dependence exactly cancels between the resummation factors of soft and beam functions, cf. \eqs{S_logNu_resum}{B_logNu_resum}.

Switching off the (\MSbar) $\mu$ evolution by setting $\mu_B=\mu_S=\mu$ we can easily identify the exponential on the RHS of \eq{IIS} with the product of the soft and beam function resummation factors.
Concretely we have
\begin{align}
\exp\bigg[{\gamma_{S,\nu}^k}(\bb,\mu)\ln\frac{\nu}{\nu_s} +{\gamma_{B,\nu}^k}(\bb,\mu)\bigg(\ln\frac{\nu}{\nu^+_B}+\ln\frac{\nu}{\nu^-_B}\bigg)\bigg] =
\nn\\
=\bigg(\frac{\nu^+_B \nu^-_B}{\nu_S^2}\bigg)^{\frac{1}{2}\gamma_\nu^k(\bb,\mu)}
=
\bigg(\frac{\mathbf{b}^2Q^2}{4e^{-2\gamma_E}}\bigg)^{-F_{k}(\bb,\mu)}
\,.
\label{eq:resummed_RRG}
\end{align}
The last equality requires that we set $\nu_S$ and $\nu_B^\pm$ to the canonical values in \eq{defaultscales} and
\begin{align}
\label{eq:gammaNu_vs_F}
\frac{1}{2}\gamma_\nu^k(\bb,\mu)=-\gamma_{B,\nu}^k(\bb,\mu)=-F_{k}(\bb,\mu)
\,.
\end{align}
Exploiting the symmetry between the $n$-collinear and $\bn$-collinear beam functions, we can now extract the unknown coefficients iteratively from \eq{IIS} order by order in $\alpha_s$ for each combination of partons $k,i,j$.
Let us sketch two possible ways to do so:

$(i)$ We set all scales, i.e.\ $\mu_B=\mu_S=\mu$ as well as $\nu_S$ and $\nu_B^\pm$, equal to the respective canonical values in \eq{defaultscales}.
We then divide both sides of \eq{IIS} by the exponential factor in \eq{resummed_RRG}.
In this way we have removed all logarithms from \eq{IIS} and the soft and beam function coefficients on the RHS reduce to $S^i_n$ and $I_{ij}^{(n)}(z)$, $J_{gj}^{(n)}(z)$, respectively.
We can thus directly determine the missing coefficients $I_{ij}^{(2)}(z)$, $J_{gj}^{(2)}(z)$.

$(ii)$ We expand \eq{IIS} to $\ord(\alpha_s^2)$. Then the dependence on $\nu$ (and of course on $\mu_S$, $\mu_B$, $\nu_S$, $\nu_B^\pm$) on the LHS drops out. The only $\mu$-dependent logarithm on both sides of the equation is therefore $L_b$.
We now write
\begin{align}
\label{eq:separate_LQ}
 \ln \frac{\mathbf{b}^2Q^2}{4e^{-2\gamma_E}} = L_b - \ln \frac{\nu}{\omega_+} - \ln \frac{\nu}{\omega_-}  -2\ln \frac{\mu}{\nu}
\,.
\end{align}
on the RHS and plug in the soft function coefficients in \eqss{Sbspace0}{Sbspace2} on the LHS.
The separation of logarithms in \eq{separate_LQ} is motivated by the structure of the soft function and the symmetry between the matching coefficients of the $n$- and $\bn$-collinear beam functions.
In fact it reveals the physical origin of the hard scale $Q^2=\omega_+ \omega_-$ in the combined soft-collinear sector as an interplay of the rapidity related scales.
We can thus uniquely identify the complete matching kernels $\mathcal{I}_{ki}(z_n,\bb,\mu,\omega_\pm/\nu)$.

Method $(i)$ is somewhat easier to carry out, but relies on the recursion relation in \eq{Irecurrence}.
In contrast, method $(ii)$ does not require any knowledge about the logarithmic structure of the beam functions.
Instead the comparison of the result to \eq{Irecurrence} serves as a strong consistency check of both frameworks and our result for the soft function.
Another strong check is that our result for $\gamma^i_\nu$ confirms \eqref{eq:gammaNu_vs_F}.
We have independently verified that both methods work perfectly.
We present the results for the $\mathcal{I}_{ki}(z_n,\bb,\mu,\omega_\pm/\nu)$ in the next section.

Finally we emphasize again that the complete factorization of logarithms according to \eq{separate_LQ} and the freedom to choose the $\nu$ scales is an advantage of using the RRG formalism. Eventually this provides the complete set of relevant scale variations as a tool to estimate the uncertainty in the resummed cross section.

\section{TMD beam functions}
\label{sec:beam}

As described in \subsec{rel_BN} we use our soft NNLO results and \eq{IIS} to extract the NNLO TMD beam functions in the RRG scheme from \mycites{Gehrmann:2012ze,Gehrmann:2014yya}. Here we present the results.

\subsection{Results in position space}
\label{subsec:bspace_I}

In $\bb$ space we parametrize the matching kernel of the gluon beam function as ($x_\perp^2 = - \bb^2$)
\begin{align}
\sum_{i}\bigg[\frac{g_\perp^{\mu\nu}}{2}\mathcal{I}_{gi}(z,\bb,\mu,\omega_\pm/\nu) +\left(\frac{g_\perp^{\mu\nu}}{2} + \frac{x_\perp^\mu x_\perp^\nu}{\bb^2}\right)\mathcal{J}_{gi}(z,\bb,\mu,\omega_\pm/\nu)\bigg]\,.
  \label{eq:IJgluonpos}
\end{align}

We first give explicit expressions for the beam function matching coefficients for $n=0,1,2$ according to the recurrence relation in \eq{Irecurrence}.
We start at LO, i.e. $\ord(\alpha_s^0)$: 
\begin{align}
  \mathcal{I}_{ij}^{(0)}(z,\bb,\mu,\omega_\pm/\nu)&=\delta_{ij}\delta(1-z)\,,\\
  \mathcal{J}_{gi}^{(0)}(z,\bb,\mu,\omega_\pm/\nu)&=0\,.
  \label{eq:Jgi0}
\end{align}
We then proceed iteratively and find 
\begin{align}
  \mathcal{I}_{ij}^{(1)}(z,\bb,\mu,\omega_\pm/\nu)&=\Big[\Big(\Gamma_0^i\ln\frac{\nu}{\omega_\pm}+\frac{1}{2}{\gamma_B^i}_0 \Big)\delta_{ij}\delta(1-z)-2P_{ij}^{(0)}(z)\Big]L_b+I_{ij}^{(1)}(z)\,,\label{eq:I1}\\
  \mathcal{J}_{gi}^{(1)}(z,\bb,\mu,\omega_\pm/\nu)&=J_{gi}^{(1)}(z)\,,
\end{align}
and
\begin{align}
  \mathcal{I}_{ij}^{(2)}(z,\bb,\mu,\omega_\pm/\nu)&=\bigg\{\bigg[\Big(2\Gamma_0^i\ln\frac{\nu}{\omega_\pm}+{\gamma_B^i}_0\Big)^2+4\Gamma_0^i\beta_0\ln\frac{\nu}{\omega_\pm}+2\beta_0{\gamma_B^i}_0\bigg]\frac{\delta_{ij}\delta(1-z)}{8} \nn\\
  & -\bigg(2\Gamma_0^i\ln\frac{\nu}{\omega_\pm}+{\gamma_B^i}_0+\beta_0\bigg)P_{ij}^{(0)}(z)+2\sum_{k} P_{ik}^{(0)}(z) \otimes_z P_{kj}^{(0)}(z)\bigg\}L_b^2\nn\\
  &+\bigg\{\bigg[\left(\Gamma_1^i\ln\frac{\nu}{\omega_\pm}+\frac{1}{2}{\gamma_B^i}_1\right)\delta_{ij}\delta(1-z)-4P_{ij}^{(1)}(z)\bigg]
  \nn\\
  &+\bigg[\Gamma_0^i\ln\frac{\nu}{\omega_\pm}+\frac{1}{2}{\gamma_B^i}_0+\beta_0\bigg]I_{ij}^{(1)}(z)-2\sum_{k}I_{ik}^{(1)}(z) \otimes_z P_{kj}^{(0)}(z)\bigg\}L_b\nn\\
  &-\frac{1}{2}{\gamma_\nu^i}_1\delta_{ij}\delta(1-z)\ln\frac{\nu}{\omega_\pm}+I_{ij}^{(2)}(z),\\
  \mathcal{J}_{gj}^{(2)}(z,\bb,\mu,\omega_\pm/\nu)&=\bigg[\bigg(\Gamma_0^g\ln\frac{\nu}{\omega_\pm}+\frac{1}{2}{\gamma_B^g}_0+\beta_0\bigg)J_{gj}^{(1)}(z)-2\sum_{k} J_{gk}^{(1)}(z) \otimes_z P_{kj}^{(0)}(z)\bigg]L_b\nn\\
  &+J_{gj}^{(2)}(z)\label{eq:J2}\,.
\end{align}
The splitting functions $P_{ij}^{(n)}(z)$ in our notation can be found in \mycites{Gaunt:2014xga,Gaunt:2014cfa}.
The (Mellin) convolutions can be performed easily e.g.\ with the {\tt Mathematica} package {\tt MT}~\cite{Hoeschele:2013gga}.

The non-cusp anomalous dimension ${\gamma_B^i}$ can be obtained by analyzing the $\mu$ dependence of \eq{IIS}.
We find
\begin{align}
{\gamma_B^i}=-2\gamma^i-\frac{1}{2}{\gamma_S^i},
\end{align}
with $\gamma^i$ as defined in \mycite{Gehrmann:2014yya}.
This yields the following constants in the expansion according to \eq{ExpGams}
\begin{align}
  {\gamma_B^g}_0&=2\beta_0\,,\label{eq:gB1}\\
  {\gamma_B^q}_0&=6C_F\,,\\
  {\gamma_B^g}_1&=C_A^2(24\zeta(3)-8)+C_A(8\beta_0-22C_F)+6\beta_0C_F\,,\\
  {\gamma_B^q}_1&=C_F^2(48\zeta(3)-4\pi^2+3)+C_F\Big[\beta_0\left(1+\frac{4\pi^2}{3}\right)+C_A(2-24\zeta(3))\Big]
  .
\end{align}

From the procedure described in \subsec{rel_BN} we finally obtain the constants in \eqss{I1}{J2}:
\begin{align}
  I_{q'q}^{(1)}(z)&= I_{\bar{q}q}^{(1)}(z)=0 \,,\\
  I_{gi}^{(1)}(z)&= 2C_F z \,(\delta_{qi}+\delta_{\bar{q}i}) \,,\\
  I_{qi}^{(1)}(z)&= 2C_F(1-z)\,\delta_{qi}+4T_Fz(1-z)\,\delta_{gi} \,,\\
  I_{q'q}^{(2)}(z)&= I_{q'/q}^{(2)}(z,0) \,,\\
  I_{\bar{q}q}^{(2)}(z)&= I_{\bar{q}/q}^{(2)}(z,0) \,,\\
  I_{gi}^{(2)}(z)&=\frac{1}{2}\delta(1-z)
  \bigg[C_A^2 \frac{\pi^4}{36} - S^g_2\bigg] \delta_{gi} + C_FC_A 
  \frac{\pi^2}{3}z \,\delta_{qi} +I_{g/i}^{(2)}(z,0)\,,\\
  I_{qi}^{(2)}(z)&=\frac{1}{2}\delta(1-z)\bigg[C_F^2 \frac{\pi^4}{36} 
  - S^q_2\bigg] \delta_{qi}  +  C_F^2 \frac{\pi^2}{3}(1-z)\, \delta_{qi} \nn\\
  &+  2  C_FT_F \frac{\pi^2}{3} z(1-z)\, \delta_{gi}  +I_{q/i}^{(2)}(z,0) \,,\\
  J_{gi}^{(1)}(z)&=4\frac{1-z}{z}\Big[C_A\delta_{gi}+C_F (\delta_{qi}+\delta_{\bar{q}i})\Big]\,.
  \label{eq:J1}
\end{align}	
Explicit expressions for the $I_{i/j}^{(2)}(z,0)$ are given in section~4 of \mycite{Gehrmann:2014yya}.

The $J_{gi}^{(2)}$ matching constants have not been calculated in any scheme so far.
For the case of scalar and vector boson production (including their decays) the $J_{gi}^{(2)}$ are however not required for predictions with NNLL$'$ accuracy~\cite{Catani:2010pd,Gehrmann:2014yya}. The reason is that the Lorentz structure associated with the $\mathcal{J}_{gj}$ in \eq{Bg} is orthogonal to $g^{\mu\nu}$ and does not occur at LO, cf. \eq{Jgi0}.

Note that due to charge conjugation and flavor symmetry, the above expressions determine the beam functions for all partons.
Respecting flavor symmetry we only wrote $q$ and $q' \neq q$ to denote quarks of unspecified flavor. 
For two quarks of flavor $a$ and $b$ we can write e.g. $I_{q_a q_b} = \delta_{ab} I_{qq} + (1-\delta_{ab}) I_{q'q}$.
From charge conjugation symmetry we have $I_{\bar{r}\bar{s}}=I_{rs}$ and $I_{r\bar{s}}=I_{\bar{r}s}$, with $r,\,s\in\{g,q,q'\}$.
Up to two loops also $I_{\bar{q}'q}=I_{q'q}$.

\subsection{Results in momentum space}
\label{subsec:pspace}

For completeness we here give the NNLO beam function results in momentum space.
We obtain them by Fourier transforming the position space beam functions in \subsec{bspace} using \Tab{fourier2}.
Care has to be taken in the transformation of the coefficients $\mathcal{J}_{ij}$ as the second Lorentz tensor in \eq{IJgluonpos} itself depends on $\bb$.
We again express the matching coefficients in \eqs{Bq}{Bg} as a power series in $a_s$:%
\footnote{Note the factor of $(2\pi)^2$ difference to \eq{IijbspaceExp}.}
\begin{align}
\label{eq:I_coef}
  \mathcal{I}_{ij}(z,\pt,\mu,\omega_\pm/\nu)&=\sum_{n=0}^\infty \mathcal{I}_{ij}^{(n)}(z,\pt,\mu,\omega_\pm/\nu) \, a_s^n \,,		\\
  	 \mathcal{J}_{gi}(z,\pt,\mu,\omega_\pm/\nu)&=\sum_{n=0}^\infty \mathcal{J}_{gi}^{(n)}(z,\pt,\mu,\omega_\pm/\nu)\, a_s^n \,,
\end{align}
with
\begin{align}
  \mathcal{I}_{ij}^{(0)}(z,\pt,\mu,\omega_\pm/\nu)&=\delta_{ij}\delta(1-z)\plus{-1} \,,\\
  \mathcal{J}_{gi}^{(0)}(z,\pt,\mu,\omega_\pm/\nu)&=0 \,,\\
  \mathcal{I}_{ij}^{(1)}(z,\pt,\mu,\omega_\pm/\nu)&=\bigg[-\Big(2\Gamma_0^i\ln\frac{\nu}{\omega_\pm}+{\gamma_B^i}_0\Big)\delta_{ij}\delta(1-z)+4P_{ij}^{(0)}(z)\bigg]\plus{0}+I_{ij}^{(1)}(z)\plus{-1}\, ,\label{eq:I1_m}\\
  \mathcal{J}_{gi}^{(1)}(z,\pt,\mu,\omega_\pm/\nu)&=-2J_{gi}^{(1)}(z)\plus{0} \,,\\
  \mathcal{I}_{ij}^{(2)}(z,\pt,\mu,\omega_\pm/\nu)&=\bigg\{-\bigg[\frac{1}{2}\Big(2\Gamma_0^i\ln\frac{\nu}{\omega_\pm} + {\gamma_B^i}_0\Big)^2+2\Gamma_0^i\beta_0\ln\frac{\nu}{\omega_\pm}+\beta_0{\gamma_B^i}_0\bigg]\delta_{ij}\delta(1-z)\nn\\
  &+\Big(8\Gamma_0^i\ln\frac{\nu}{\omega_\pm}+4{\gamma_B^i}_0+4\beta_0\Big)P_{ij}^{(0)}(z)-8\sum_{k} P_{ik}^{(0)}(z)\otimes_z P_{kj}^{(0)}(z) \bigg\}\plus{1}\nn\\
  &+\bigg[-\Big(2\Gamma_1^i\ln\frac{\nu}{\omega_\pm}+{\gamma_B^i}_1\Big)\delta_{ij}\delta(1-z)+8P_{ij}^{(1)}(z)\nn\\
  &-\Big(2\Gamma_0^i\ln\frac{\nu}{\omega_\pm}+{\gamma_B^i}_0+2\beta_0\Big)I_{ij}^{(1)}(z)+4\sum_{k}{ I_{ik}^{(1)}(z) \otimes_z P_{kj}^{(0)}(z) }\bigg]\plus{0}\nn\\
  &+\bigg[-\frac{1}{2}{\gamma_\nu^i}_1\delta_{ij}\delta(1-z)\ln\frac{\nu}{\omega_\pm}+I_{ij}^{(2)}(z)\bigg]\plus{-1} \,,\\
  \mathcal{J}_{gi}^{(2)}(z,\pt,\mu,\omega_\pm/\nu)&=\bigg[-\Big(2\Gamma_0^g\ln\frac{\nu}{\omega_\pm}+{\gamma_B^g}_0+2\beta_0\Big)J_{gi}^{(1)}(z)\nn\\
  &+4\sum_{k}{ J_{gk}^{(1)}(z) \otimes_z P_{ki}^{(0)}(z) }\bigg]\left(\plus{0}+\plus{1}\right)-2J_{g { i}}^{(2)}(z)\plus{0}\label{eq:J2_m} \,.
\end{align}
The ${\gamma_B^i}_n$, $I_{ij}^{(n)}(z)$ and $J_{gi}^{(n)}(z)$ are the same as in the position space results and are written down in \eqss{gB1}{J1}.
The one-loop results are in agreement with \mycites{Chiu:2012ir,Ritzmann:2014mka}.

\section{Conclusions}
\label{sec:conclusions}

We have calculated the TMD soft function in the RRG scheme of \mycites{Chiu:2011qc,Chiu:2012ir} to NNLO.
To regularize rapidity divergences we have employed the $\eta$-regulator~\cite{Chiu:2011qc,Chiu:2012ir}.
We have explicitly demonstrated that this regulator preserves non-Abelian exponentiation of the TMD soft function at two loops. This represents a valuable consistency and practicability check of the regularization method.
We present the new result in momentum ($\pt$) as well as in position ($\bb$) space in \eqs{S2}{Sbspace2}, respectively.
We also obtain the soft two-loop anomalous dimension for the RGE and RRGE.
The corresponding expressions in \eqs{gammaconst2}{gammaconst3} exhibit an interesting similarity.

Based on the equality of cross section predictions and with the two-loop soft function at hand we have furthermore extracted the NNLO TMD beam functions in the RRG scheme from the results of \mycites{Gehrmann:2012ze,Gehrmann:2014yya}.
We have checked that our expressions for the soft and beam functions match the logarithmic structure predicted by recursion relations we have derived from the corresponding RGEs and RRGEs.
The results for the beam function matching kernels and anomalous dimension are collected in \sec{beam}.
All expressions for the soft and beam function coefficients are also available in electronic form upon request to the authors.

Our results represent the universal ingredients in the SCET factorization theorem for the peak region of the transverse momentum distribution of a color-neutral final state at the LHC with RRG resummation at NNLL$'$.
The perturbative uncertainties of such resummed cross sections can be systematically studied by independent variations of the different involved renormalization/factorization scales associated with the RG as well as the RRG.
The phenomenological analysis at NNLL$'$(+NNLO) of the transverse momentum spectra for processes like Drell-Yan or Higgs production is left to future work.

\begin{acknowledgments}
We like to thank Frank Tackmann and Markus Ebert for many useful discussions and reading the manuscript.
Special thanks to Frank Tackmann, who suggested this project.
TL and MS thank the Mainz Institute for Theoretical Physics (MITP) for hospitality during part of this work.
JO thanks DESY for hospitality and support.
This work was supported in parts by the DFG Emmy-Noether Grant No. TA 867/1-1, the DFG Grant SFB 676-B11 and the Swedish Science Research Council Grant 621-2011-5333.
All diagrams were drawn with JaxoDraw \cite{Binosi:2008ig}.
\end{acknowledgments}

\appendix

\section{Plus distributions}
\label{app:plusDistributions}

When expanding the bare expressions in $\eta$ and $\e$, we make use of the distributional identity for $\mu^2, \pt^2>0$,
\begin{align}\label{eq:plusidentity}
\frac{1}{2 \pi \mu^2} \bigg(\frac{\mu^2}{\pt^2}\bigg)^{\!1+\alpha} \,=\,
-\frac{\plus{-1}(\pt)}{2 \alpha} + \sum_{n=0}^\infty\frac{\alpha^n}{n!}\plus{n}(\pt,\mu) \,,
\end{align}
where we define the $\plus{n}(\pt,\mu)$ for $n \ge 0$ in terms of the usual plus distributions%
\footnote{Here we introduce the superscript $T$ to distinguish the $\pt$-dependent plus distributions from the notation $\mathcal{L}_n(x)=[\theta(x)/x \ln^n x]_+$ as introduced in \mycite{Ligeti:2008ac}. It holds $\plus{n}(\pt,\mu)=\frac{(-1)^n}{2\pi\mu^2}\mathcal{L}_n(\pt^2/\mu^2)$ for $n\geq0$.}
\begin{align}\label{eq:plusDef1}
\plus{n}(\pt,\mu)\equiv 
\frac{1}{2\pi\mu^2}\left[\frac{\mu^2}{\pt^2}\ln^n\left(\frac{\mu^2}{\pt^2}\right)\right]_+
\,.
\end{align}
and 
\begin{align}\label{eq:plusDef2}
\plus{-1}(\pt)\equiv \frac1\pi \delta(\pt^2) = \delta^\two(\pt)\,.
\end{align}
For an equivalent definition of the $\plus{n}(\pt,\mu)$ and more details about their properties and generalizations we refer to \mycite{Chiu:2012ir}.
By definition and with our choice of boundary condition%
\footnote{This corresponds to $\lambda=1$ in \mycite{Chiu:2012ir}.}
we have,
\begin{align}
\int_{D_\mu} \!\! \df^2 p_\perp\; \plus{n}(\pt,\mu) = 0\,\quad \forall n \ge 0\,, \qquad 
\int_{D_\kappa} \!\! \df^2 p_\perp\; \plus{-1}(\pt,\mu) = 1
\,,
\end{align}
where $D_\kappa=\{\pt: |\pt|\le \kappa\}$ denotes a disc in $\pt$-space around the origin with radius $\kappa$,
which is set to $\mu$ in the first and an arbitrary positive value in the second equation.

When dealing with the RGE structure of the theory, it is often convenient to work in position space where the plus distributions become ordinary logarithms and convolutions turn into ordinary products.
In the following we give explicit expressions for the relevant Fourier transformations as well as for the convolutions of the $\plus{n}$.

\subsection*{Fourier transformations}

In \Tab{fourier} we give the results of Fourier transforming the $\plus{n}$ with $n\le4$ to $\mathbf b$ space in terms of $L_b=\ln\left(\mathbf{b}^2\mu^2e^{2\gamma_E}/4\right)$.
The inverse Fourier transformations are collected in \Tab{fourier2}.

\begin{table}[h!]
\begin{center}
\begin{tabular}{|c|c|}
\hline
$\mathbf{p}$-space 	&	$\mathbf{b}$-space\\
\hline
$\plus{-1}$	&	$\frac{1}{(2\pi)^2}$\\
$\plus{0}$	&	$-\frac{1}{8\pi^2}L_b$\\
$\plus{1}$	&	$-\frac{1}{16\pi^2}L_b^2$\\
$\plus{2}$	&	$-\frac{1}{24\pi^2}\left[L_b^3+4\zeta(3)\right]$\\
$\plus{3}$	&	$-\frac{1}{32\pi^2}\left[L_b^4+16\zeta(3)L_b\right]$\\
$\plus{4}$	&	$-\frac{1}{40\pi^2}\left[L_b^5+40\zeta(3)L_b+48\zeta(5)\right]$\\
\hline
\end{tabular}
\caption{Fourier transforms of plus distributions $\plus{n}$ in terms of $L_b=\ln\left(\mathbf{b}^2\mu^2e^{2\gamma_E}/4\right)$.}
\label{tab:fourier}
\end{center}
\end{table}

\begin{table}[h!]
\begin{center}
\begin{tabular}{|c|c|}
\hline
$\mathbf{b}$-space 	&	$\mathbf{p}$-space\\
\hline
$1$	&	$(2\pi)^2\plus{-1}$ \\
$L_b$	&	$-8\pi^2\plus{0}$\\
$L_b^2$	&	$-16\pi^2\plus{1}$\\
$L_b^3$	&	$-8\pi^2\left[3\plus{2}+2\zeta(3)\plus{-1}\right]$\\
$L_b^4$	&	$-32\pi^2\left[\plus{3}-4\zeta(3)\plus{0}\right]$\\
\hline
\end{tabular}
\caption{Fourier transforms of $L_b^n=[\ln\left(\mathbf{b}^2\mu^2e^{2\gamma_E}/4\right)]^n$ to momentum space in terms of plus distributions.}
\label{tab:fourier2}
\end{center}
\end{table}

\subsection*{Convolutions}
Here we list the convolutions of plus distributions relevant for this work. They can be derived from expanding eq.~(F.23) in \mycite{Chiu:2012ir}:
\begin{align}
\plus{0}\otimes_\perp\plus{0}&=-\frac{1}{4\pi^2}\plus{1},\\
\plus{0}\otimes_\perp\plus{1}&=-\frac{3}{16\pi^2}\plus{2}-\frac{\zeta(3)}{8\pi^2}\plus{-1},\\
\plus{0}\otimes_\perp\plus{2}&=-\frac{1}{6\pi^2}\plus{3}+\frac{\zeta(3)}{2\pi^2}\plus{0},\\
\plus{1}\otimes_\perp\plus{1}&=-\frac{1}{8\pi^2}\plus{3}+\frac{\zeta(3)}{2\pi^2}\plus{0},\\
\plus{1}\otimes_\perp\plus{2}&=-\frac{5}{48\pi^2}\plus{4}+\frac{3\zeta(3)}{2\pi^2}\plus{1}-\frac{\zeta(5)}{2\pi^2}\plus{-1}
\,.
\end{align}

\section{Expansion of hypergeometric functions}
\label{app:expansion}
For certain sets, $\{a_i\}$, of parameters, the solution of $I^{RR}(\{a_i\})$ contains a hypergeometric function $_3F_2$, see \eqss{Isol1}{Isol3}.
In most cases we encounter, one of the first three indices of the hypergeometric function is a non-positive integer, while the last two indices are no integers. These hypergeometric functions then trivially reduce to rational functions in the regulators.
For the three non-trivial cases, we used the {\tt Mathematica} package {\tt HypExp}~\cite{Huber:2005yg,Huber:2007dx} to expand the hypergeometric function and obtained
\begin{align}
I^{RR}(1,1,0,0,0,0)&=\frac{1}{\epsilon^2}+\frac{\pi^2}{3}+4\zeta(3)\epsilon+\frac{11\pi^4}{90}\epsilon^2+\ordo{\e^3} \nonumber\\
&\hspace{-1cm}+\eta\left[\frac{\pi^2}{6\epsilon}+\frac{17\pi^4}{360}\epsilon+\ordo{\e^2}\right]+\eta^2\left[-\frac{\zeta(3)}{2\e}+\frac{\pi^4}{72}+\ordo{\e^2}\right]+\ordo{\eta^3},\\
I^{RR}(0,1,1,0,0,0)&=\frac{1}{\eta}\left[-\frac{4}{\epsilon}+8\zeta(3)\epsilon^2+\frac{2\pi^4}{15}\epsilon^3+\ordo{\e^4}\right]\nonumber\\
&-\frac{\pi^2}{3}+2\zeta(3)\epsilon-\frac{\pi^4}{45}\epsilon^2+\ordo{\e^3}+\eta\left[\zeta(3)-\frac{11\pi^4}{360}\epsilon+\ordo{\e^2}\right]\nonumber\\
&+\eta^2\left[-\frac{\pi^4}{180}+\ordo{\e}\right]+\ordo{\eta^3},\\
I^{RR}(0,1,1,0,0,1)&=\frac{1}{\eta}\left[-\frac{4}{\e}+8\zeta(3)\e^2+\frac{2\pi^4}{15}\e^3+\ordo{\e^4}\right]\nonumber\\
&+\frac{2}{\e^2}-\frac{2\pi^2}{3}-4\zeta(3)\e-\frac{\pi^4}{9}\e^2+\ordo{\e^3}\nonumber\\
	&\hspace{-1cm}+\eta\left[\frac{\pi^2}{6\e}+2\zeta(3)-\frac{\pi^4}{30}\e+\ordo{\e^2}\right]+\eta^2\left[-\frac{\zeta(3)}{2\e}-\frac{\pi^4}{144}+\ordo{\e}\right]+\ordo{\eta^3}.
\end{align}
The $_3F_2$ appearing in the last integral could not be expanded directly with HypExp.
Using the integral representation of $_3F_2$, expanding the $_2F_1$ in the integrand, and then integrating all the terms separately leads to the stated result.

\bibliographystyle{jhep}
\bibliography{pT}

\end{document}